\documentclass[iop,apj,numberedappendix,appendixfloats]{emulateapj}

\usepackage{amsmath}
\usepackage{multirow}

\bibpunct[,]{(}{)}{;}{a}{}{,}

\newcommand{\Omegam}{\ensuremath{\Omega_{\mathrm{m}}}}
\newcommand{\OmegaL}{\ensuremath{\Omega_{\Lambda}}}

\newcommand{\hmpc}{\; h^{-1}\mathrm{Mpc}}

\newcommand{\hMsun}{\; h^{-1}M_{\odot}}
\newcommand{\kms}{{\; {\rm km}\,{\rm s}^{-1}}}
\newcommand{\hunit}{\kms {\mathrm{Mpc}}^{-1}}

\newcommand{\hvolm}{\; h^{-3}{\mathrm{Mpc}}^{3}}
\newcommand{\hvolg}{\; h^{-3}{\mathrm{Gpc}}^{3}}

\newcommand{\mpart}{m_\mathrm{part}}

\newcommand{\lcdm}{\ensuremath{\Lambda\mathrm{CDM}}}
\newcommand{\Lstar}{\ensuremath{L_{\ast}}}

\newcommand{\ntropy}{{\it N}tropy}
\newcommand{\nbody}{$N$-body}
\newcommand{\npoint}{$n$-point}
\newcommand{\pimax}{\pi_\mathrm{max}}

\newcommand{\band}[2]{\ensuremath{^{#1}\!{#2}}}
\newcommand{\Mr}{M_{\band{0.1}{r}}}

\newcommand{\Hunits}{{\,{\rm km}\,{\rm s}^{-1}\,{\rm Mpc}^{-1}}}

\begin{document}

\title{Three-Point Correlation Functions of SDSS Galaxies: Luminosity and Color Dependence in 
Redshift and Projected Space}

\author{
  Cameron~K.~McBride\altaffilmark{1,2},
  Andrew~J.~Connolly\altaffilmark{3}, 
  Jeffrey~P.~Gardner\altaffilmark{4}, 
  Ryan~Scranton\altaffilmark{5}, 
  Jeffrey~A.~Newman\altaffilmark{1}, 
  Rom\'an~Scoccimarro\altaffilmark{6}, 
  Idit~Zehavi\altaffilmark{7}, 
  Donald~P.~Schneider\altaffilmark{8}
}
\submitted{Accepted to ApJ: 14 October 2010}
\email{cameron.mcbride@vanderbilt.edu}


\altaffiltext{1}{Department of Physics \& Astronomy, University of Pittsburgh, Pittsburgh, PA 15260} 
\altaffiltext{2}{Department of Physics and Astronomy, Vanderbilt University, Nashville, TN 37235}
\altaffiltext{3}{Department of Astronomy, University of Washington, Seattle, WA 98195-1580}
\altaffiltext{4}{Department of Physics, University of Washington, Seattle, WA 98195-1560}
\altaffiltext{5}{Department of Physics, University of California, Davis, CA 95616}
\altaffiltext{6}{Center for Cosmology and Particle Physics, New York University, 4 Washington Place, New York, NY 10003}
\altaffiltext{7}{Department of Astronomy, Case Western Reserve University, Cleveland, OH 44106}
\altaffiltext{8}{Department of Astronomy \& Astrophysics, Pennsylvania State University, University Park, PA 16802}


\begin{abstract}

The three-point correlation function (3PCF) provides an important view into the 
clustering of galaxies that is not available to its lower order cousin, the 
two-point correlation function (2PCF).  Higher order statistics, such as the 3PCF, 
are necessary to probe the non-Gaussian structure and shape information expected 
in these distributions.
We measure the clustering of spectroscopic galaxies in the Main Galaxy Sample of the Sloan
Digital Sky Survey (SDSS), focusing on the shape or \emph{configuration} dependence of the
reduced 3PCF in both redshift and projected space.  This work constitutes the largest 
number of galaxies ever used to investigate the reduced 3PCF, using over $220\,000$ galaxies 
in three volume-limited samples. 
We find significant configuration dependence of the reduced 3PCF at $3-27 \hmpc$, in 
agreement with \lcdm\ predictions and in disagreement with the hierarchical ansatz.  Below 
$6 \hmpc$, the redshift space reduced 3PCF shows a smaller amplitude and weak configuration 
dependence in comparison with projected measurements suggesting that redshift distortions, 
and not galaxy bias, can make the reduced 3PCF appear consistent with the hierarchical 
ansatz. 
The reduced 3PCF shows a weaker dependence on luminosity than the 2PCF, with no 
significant dependence on scales above $9 \hmpc$. On scales less than $9 \hmpc$, the 
reduced 3PCF appears more affected by galaxy color than luminosity. 
We demonstrate the extreme sensitivity of the 3PCF to systematic effects such as sky
completeness and binning scheme, along with the difficulty of resolving the errors. 
Some comparable analyses make assumptions that do not consistently account for these 
effects.

\end{abstract}

\section{Introduction}
\label{s:intro}

In the current paradigm of structure formation, an almost uniform distribution of mass at 
early times in the universe evolved through gravitational instability into the irregular 
and complex distribution that galaxies occupy today.  
Gravitational dynamics are sensitive to cosmology and depend on the spatial curvature of
the universe as well as the nature of its contents.  A combination of many
observations including both constraints from large-scale structure 
\citep[LSS;][]{tegmark:04_ps,cole:05, eisenstein:05,tegmark:06,sanchez:09} and the 
cosmic microwave background \citep[CMB; ][]{spergel:07} support a standard model in accordance 
with current observations \citep[see recent constraints in][]{komatsu:09}. These observations 
suggest a critically dense (spatially flat) universe, consisting of a small amount of baryonic matter, 
several times more mass in cold dark matter (CDM), and well over the majority 
of the current energy density in some form of dark energy ($\Lambda$).  This concordance model, 
referred to as $\lcdm$, forms the basis of predicting LSS and the framework underlying 
statistical studies of galaxy distributions.

Modern galaxy surveys, such as the Sloan Digital Sky Survey \citep[SDSS; ][]{york:00,stoughton:02}, 
provide a wealth of information about large-scale structure, galaxy formation, galaxy evolution and cosmology. 
The main workhorse for statistical analyses has been the \npoint\ correlation functions
\citep{peebles:80}.  The most common clustering measurements use the two-point correlation 
function (2PCF) or its analog in Fourier space, the power spectrum.  Results from measurements of 
these statistics using observational datasets have been able to distinguish subtle differences 
in theoretical models of how galaxies occupy dark matter halos \citep[e.g. departures from power 
law clustering, ][]{zehavi:04}, a comprehensive study of the effects of luminosity and color 
on galaxy clustering \citep{zehavi:05,zehavi:10}, as well as providing precise measurements of 
cosmological phenomena to better understand the nature of dark energy \citep[e.g. baryon 
acoustic oscillations, ][]{eisenstein:05}. 

If the galaxy distribution was entirely Gaussian, the 2PCF would provide a complete 
description of galaxy clustering.  Although analyses of the CMB suggest that the
primordial mass fluctuations in our universe appear extremely Gaussian \citep{komatsu:09}, 
we expect gravitational collapse to produce non-Gaussian signatures in the galaxy distribution 
that we measure today \citep{lss_review}.  As such, the 2PCF provides only a partial view 
of the full distribution and cannot sufficiently probe non-Gaussian signals.

To investigate non-Gaussian structure, as well as shape information in these
distributions, we require higher order clustering statistics.  In the hierarchy of 
\npoint\ correlation functions, the 3PCF is the lowest order statistic to
provide information on shape \citep{peebles:80}.  For example, this enables probes of 
the triaxial nature of halos and extended filaments within the ``cosmic web''.  
Measurements of higher order moments allow a more complete picture of the galaxy 
distribution, breaking model degeneracies describing cosmology and galaxy bias 
\citep{zheng:07,kulkarni:07}.  

The statistical strength of information in higher order moments of clustering 
can potentially rival that of the two-point statistics \citep{sefusatti:05}.  As a 
complement to bispectrum (Fourier transform of the 3PCF) and redshift-space 
analyses, we can use the projected 3PCF to sidestep redshift distortions in 
observational data by integrating out the effect of peculiar velocities from 
the density field.  

The additional information contained in higher order moments comes at a price. Their
increased complexity make the measurements, modeling and interpretation difficult.
Theoretically, non-linear contributions have significant non-trivial implications.  Their
calculation gets computationally challenging and efficient algorithms become critically
important. They require larger and cleaner galaxy samples as they show more sensitivity 
to observational systematics than the 2PCF.  As it was recently described: ``the overlap
between well understood theory and reliable measurements is in fact disquietingly small''
\citep{szapudi:09}.  This work attempts to increase this overlap by leveraging the massive 
dataset available from the SDSS.

Previous work has estimated the 3PCF from modern galaxy redshift surveys, including 
work on the the two-degree field galaxy redshift survey \citep{jing:04,wang:04,gaztanaga:05} 
and results from earlier SDSS data \citep{kayo:04,nichol:06,kulkarni:07}. 
Fourier space analogs or related higher order statistics have also been measured for 
these datasets \citep{verde:02,pan:05,hikage:05,nishimichi:07}.

This work is the first of two papers analyzing the reduced 3PCF on SDSS galaxy samples, 
where this work focuses on details of the measurements as well as the dependence of 
clustering from varying the galaxy sample's luminosity and color.  The second paper 
(McBride, et al. 2010) utilizes the configuration dependence to constrain non-linear 
galaxy-mass bias parameters in the local bias model \citep{fry:93}.  Throughout our study, 
we assume a standard flat $\lcdm$ cosmology where $\Omegam = 0.3$, $\OmegaL = 0.7 $, 
and $H_o = h \, 100 \hunit$ .

This paper is organized as follows. We discuss the SDSS data and the simulations in 
\S\ref{s:data}.  We review the relevant theory and methods of our analysis in 
\S\ref{s:methods}.  We introduce our measurements in \S\ref{s:results}, which include 
measurements of the 2PCF in \S\ref{ss:2pcf}, the equilateral reduced 3PCF in 
\S\ref{ss:Qeq}, and configuration dependence of the reduced 3PCF in \S\ref{ss:Q_conf}.  We 
resolve the covariance of our measurements in \S\ref{ss:covar} and investigate the effects 
of large overdense structures, such as the Sloan Great Wall, in \S\ref{ss:sstruct}. We 
discuss our results and compare to previous studies in \S\ref{s:discussion}.  Finally, we 
review our findings in \S\ref{s:summary}.

\section{Data}
\label{s:data}

\subsection{SDSS Galaxy Samples}
\label{s:sdss}

\begin{table*}
  \centering
  \begin{tabular}{lcccccc}
    \hline
    \multicolumn{7}{c}{\bfseries Specifics of SDSS galaxy samples}  \\
    Name & \multirow{2}{*}{Magnitude} & \multirow{2}{*}{Redshift} & Volume        & Number of &  Blue / Red & Density                \\
         &                            &                           & $\hvolg   $   & Galaxies  &  Galaxies   & $10^{-3} \hvolm$ \\
    \hline
    \hline
    BRIGHT        &         $M_{r} < -21.5$ & $0.010 $ - $0.210 $ & $0.1390$ & $ 37\,875$ &   ---               & $0.272$  \\
    LSTAR         & $-21.5 < M_{r} < -20.5$ & $0.053 $ - $0.138 $ & $0.0391$ & $106\,823$ & $46\,574 / 60\,249$ & $2.732$  \\
    FAINT         & $-20.5 < M_{r} < -19.5$ & $0.034 $ - $0.086 $ & $0.0098$ & $ 76\,808$ & $39\,360 / 37\,448$ & $7.849$  \\
    \hline
  \end{tabular}
  \caption[Volume-limited galaxy catalogs from the SDSS DR6]{ 
    The redshift limits, volume, total number of galaxies, number per color sample, and 
    completeness corrected number density are shown for the three galaxy samples 
    constructed from the SDSS DR6 spectroscopic catalog.  We selected these by cuts in 
    redshift, $z$, and corrected (K-correction and passive evolution) absolute $r$-band 
    magnitude, $M_r$, to create volume-limited selections.  These samples represent 
    $221\,506$ unique galaxies.
  } \label{t:gal_samples}
\end{table*}

The SDSS \citep{york:00,stoughton:02} employs a dedicated 2.5 meter telescope 
\citep{gunn:98,gunn:06} at Apache Point Observatory in New Mexico. Nearly a quarter of the 
sky was imaged in five bandpasses \citep{fukugita:96,smith:02}, reduced by a processing 
pipeline \citep{lupton:01}, and calibrated for accurate astrometry \citep{pier:03} and 
photometry \citep{hogg:01,ivezic:04,tucker:06,padmanabhan:08}.

For our analysis, we use spectroscopic galaxy data defined as the Main galaxy sample 
\citep{strauss:02}.  The algorithm which defines this selection targets about 90 galaxies 
per square degree that turn out to have a median redshift of 0.104, a high completeness, 
and an accurate statistical separation of stars and galaxies which prevents stellar 
contaminants in the galaxy samples \citep{strauss:02}.  A fiber based spectrograph 
observes the targets using an adaptive tiling algorithm \citep{blanton:03}.

The SDSS galaxy samples are made more readily available via the New York University Value-Added 
Galaxy Catalog \citep[NYU-VAGC;][]{vagc}.  This catalog provides detailed characterizations 
of the sample geometry and completeness as well as correcting for known systematics that 
are pertinent to large-scale structure analyses.  These include $K$-corrections, passive 
evolution corrections, and ``fiber collision corrections'' that account for missing spectra 
due to galaxy pairs that are closer than fibers can be positioned on the sky ($55\arcsec$). 
We conduct our analysis of clustering measurements using galaxies from DR6 \citep{sdss_dr6}. 

The NYU-VAGC galaxy catalog corresponding to DR6 contains $\sim\!470\,000$ galaxies
covering $6377$ square degrees of unmasked area (we neglect regions around bright stars).
We select volume-limited sub-samples from this flux-limited parent catalog to analyze 
well-defined samples, which still contain a large number of galaxies ($221\,500$ unique 
galaxies in three samples).  We do not analyze a flux-limited sample.  Using volume-limited 
samples prevents systematic effects on clustering measurements from the inaccuracies in the radial 
selection function, and allows a cleaner association of clustering differences to properties 
of the galaxy sample, such as luminosity and color.

\begin{figure}
\centerline{\includegraphics[angle=0, width=1.0\linewidth]{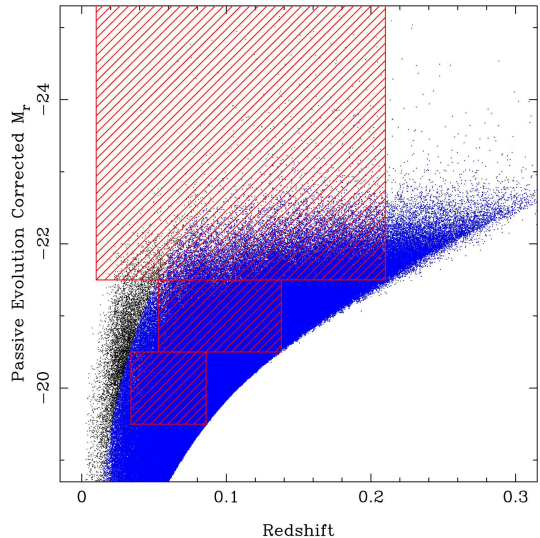}}
\caption[Volume Limited Selection for SDSS Samples]{
  This figure depicts the selection of three volume-limited samples from the SDSS DR6 
  release detailed in Table~\ref{t:gal_samples} and shown as shaded boxes corresponding to BRIGHT, 
  LSTAR, and FAINT (from top to bottom). Each point represents a galaxy, and the blue and black colors
  depict the two overlapping parent catalogs (a single galaxy can be in both catalogs). The black points 
  correspond to NYU-VAGC \texttt{bright} which does not include a limit on a galaxy's bright apparent 
  magnitude.  The blue points, overplotted on the black, correspond to NYU-VAGC \texttt{safe} and 
  include a bright cut to the apparent magnitude \citep[see][]{vagc}.  We select redshift limits 
  where galaxies of all included magnitudes can be seen at both the inner and outer boundaries.  
} \label{f:vl_sample} 
\end{figure}

We construct the volume-limited samples by using corrected absolute $r$-band magnitude as a 
function of redshift.  Our absolute $r$-band magnitudes use the NYU-VAGC 
convention: absolute magnitudes represent values at $z=0.1$ \citep[see details in][]{vagc}. 
These are often written as $\Mr$, which we simplify to $M_r$.  Basically, passive evolution 
and $K$-corrections use the median redshift of the flux-limited SDSS Main sample to 
minimize uncertainties in their empirical determination. 
Formally, $M_r = M_r - 5 \log h$ but our unit convention sets $h = 1$ making the last term 
unnecessary.  We define a sub-sample of objects by selecting bounding redshifts which correspond 
to a specific luminosity range.  We define three samples: a BRIGHT sample with $M_r < -21.5$, 
LSTAR with $-21.5 < M_r < -20.5$, and FAINT with $-20.5 < M_r < -19.5$.  We show the selection of 
these samples in Figure~\ref{f:vl_sample}. We tabulate their properties, such as the redshift 
range, volume, number of objects, and completeness corrected number density in 
Table~\ref{t:gal_samples}.

Of particular concern might be the ``fiber collision corrections'' in the galaxy data.  
The correction used in the NYU-VAGC assigns the redshift of the nearest angular neighbor to 
a galaxy that does not have a redshift, reducing the number of collisions from $\sim 7\%$ to 
less than $\sim 4\%$. This correction was investigated using galaxy mocks in the 2PCF 
\citep{zehavi:02,zehavi:05}, and concluded that it works incredibly well for separations 
above the collision scale with only a small residual effect for redshift space quantities on 
all scales.  We do not expect this correction to be appropriate for smaller scales. However, the 
fiber collision correction does not significantly affect our analysis as we focus on scales 
above $3 \hmpc$. The collision scale at the largest redshift in our galaxy sample remains 
well below this value, as $55\arcsec$ corresponds to a separation of $\sim 0.13 \hmpc$ at 
$z=0.21$.

Galaxy clustering measurements for SDSS have been shown to vary strongly with color 
\citep{zehavi:05,zehavi:10}.  We investigate the color dependence on two of our samples 
(LSTAR and FAINT) by defining a simple red-blue split based on the $g-r$ color.  There is 
a strong bi-modality between red and blue sub-samples that varies with absolute magnitude, 
as shown in \cite{baldry:04}.  We adopt a simple linearly sloped color cut that depends on 
absolute $r$-band magnitude to account for the bi-modal distribution which separates the 
red sequence from the blue cloud, as described in \cite{zehavi:05}.  Specifically, the 
$g-r$ color limit is 
\begin{equation} \label{eq:colorcut}
  (g - r)_{lim} = 0.21 - 0.03 M_r \; .
\end{equation}
Radial distances and absolute magnitudes are calculated using a flat \lcdm\ cosmology with 
($ \Omegam = 0.3 \, , \, \OmegaL = 0.7 $).

\subsection{Hubble Volume Simulation}
\label{ss:hvs}

We compare the SDSS galaxy distribution with the structure in
cosmological \nbody\ simulations. We use the \emph{Hubble Volume} (HV)
simulations \citep{colberg:00,evrard:02} that were completed by the the Virgo
Consortium.  We choose the simulation with \lcdm\ cosmology: 
($\Omegam = 0.3$, $\OmegaL = 0.7$, $H_o = 70 \Hunits$, $\sigma_8 = 0.9$).  
This HV simulation consists of $1000^3$ particles in a box of $(3000 \hmpc)^3$ 
volume which evolves particles of mass $\mpart = 2.2 \times 10^{12} \hMsun$ 
from an initial redshift of $z_{init} = 35$ to the current epoch using a Plummer 
softened gravitational potential where the softening length is $0.1 \hmpc$.
To better compare with observational galaxy samples, we utilize their ``light-cone''
realization.  The light-cone output of the HV \lcdm\ simulation was kindly provided
by Gus Evrard and J\"{o}rg Colberg.

We treat the particles of the HV simulation like a potential observation and
apply observational constraints to their distribution.  We filter particles
to match the same angular footprint of the SDSS geometry and radial distance to 
exactly reproduce the volume of the corresponding galaxy sample.  We apply 
redshift distortions by using the peculiar velocity of the DM particle.

Finally, we randomly downsample the number of dark matter particles to make the
computational time of the analysis more manageable.  To verify this downsampling 
does not introduce a systematic bias, we perform a few measurements on several downsampled 
realizations and compare with the full DM distribution. We find minimal differences 
that are within expectations for Poisson sampling.  

\section{Theory \& Methods}
\label{s:methods}

The \npoint\ correlation functions provide a statistical description of 
LSS and remain a standard tool for quantifying the structure of the mass field 
and galaxy distribution \citep{peebles:80}.  We summarize the basics relevant to this 
analysis below.

We can define density as a function of position, $\rho(\vec{x})$, that has an 
average density of $\bar{\rho}$.  We define the fractional overdensity about the 
mean at a local point as 
\begin{equation} \label{eq:overdensity}
  \delta(\vec{x}) = \frac{\rho(\vec{x})}{\bar{\rho}} - 1 \; .
\end{equation}
Casting the density in terms of the overdensity effectively removes the first moment of the 
$\delta$ field, i.e. $\langle \delta(\vec{x}) \rangle = 0$, where the $\langle \rangle$ denotes 
an ensemble average.  The \emph{two-point correlation function} (2PCF) can be defined in terms of 
$\delta$ values characterized by the separation of two positions, $r_{12} = |\vec{x}_1 - \vec{x}_2|$, 
by assuming homogeneity and isotropy, and we write the 2PCF as
\begin{equation} \label{eq:2pcf_delta}
  \xi(r_{12}) = \langle \delta(\vec{x}_1) \delta(\vec{x}_2) \rangle \; .
\end{equation}

A \emph{Gaussian} field refers to any distribution that is fully described (statistically) by only
its first and second moments (e.g. a mean and variance).  For the $\delta$ field, the mean is zero
and $\xi(r)$ successfully describes all clustering properties of a Gaussian field.  Any other possible 
distribution is then \emph{non-Gaussian} and can have non-trivial higher order moments (i.e. non-zero 
higher order correlation functions).  Higher order functions can be similarly 
defined in terms of overdensity fluctuations, where the three-point correlation function (3PCF) is given by
\begin{equation} \label{eq:3pcf_delta}
  \zeta(r_{12}, r_{23}, r_{31}) = \langle \delta(\vec{x}_1) \delta(\vec{x}_2) \delta(\vec{x}_3) \rangle \; .
\end{equation}
Instead of a single dependent variable, such as $r_{12}$ in $\xi(r_{12})$, we see the 3PCF relies on 
three separations necessary to parameterize triplets.  Further higher order correlation functions 
(greater than $n = 3$) require even more variables and quickly result in a ``combinatorial explosion'' 
of parameters \citep{szapudi:09}. 

The \emph{hierarchical ansatz} posits that the 3PCF can be estimated by a cyclic combination 
of respective 2PCFs: 
\begin{equation} \label{eq:ha}
  \zeta(r_{12}, r_{23}, r_{31}) \approx Q \left[ \xi_{12} \xi_{23} + \xi_{23} \xi_{31} + \xi_{31} \xi_{12} \right],
\end{equation}
where we have simplified notation with $\xi_{12} = \xi(r_{12})$, and $Q$ denotes a scaling constant 
to adjust the amplitude.  Initial measurements of 3PCF using angular surveys suggested that the 
hierarchical ansatz held at small scales with $Q \approx 1.3$ \citep{peebles:80}. 

What was originally called the hierarchical amplitude ($Q$) can be rewritten as a function, specifically
\begin{equation} \label{eq:Q}
  Q(r_{12}, r_{23}, r_{31}) = \frac{  \zeta(r_{12}, r_{23}, r_{31}) }
				   {\xi_{12} \xi_{23} + \xi_{23} \xi_{31} + \xi_{31} \xi_{12} }\; .
\end{equation}
This definition provides a useful normalization, and $Q(r_{12}, r_{23}, r_{31})$ is commonly referred to 
as the normalized or \emph{reduced} 3PCF.  As long as the the 2PCF remains well above zero, i.e. 
the denominator in \eqref{eq:Q}, the value of the function $Q$ roughly equals unity regardless of scale.
This functional form was later justified by gravitational perturbation theory, as the evolution of the 
3PCF depends on quadratic terms in the equations of motion encapsulated in the square of the 2PCF 
\citep{lss_review}.  An additional benefit of utilizing such a ``ratio statistic'': we expect $Q$ to 
be insensitive to both time and cosmology.  To leading order, $Q$ only depends on the spectral index 
and triangle configuration \citep[see Figure~9 in][]{lss_review}.

\subsection{Redshift Distortions and Projected Correlation Functions}

Redshift distortions destroy the isotropy of the galaxy distribution.  They arise from 
our inability to disentangle the peculiar (dynamical) velocity of a galaxy's motion from 
the Hubble flow when determining the radial distance via the spectroscopic redshift.  
We characterize the correlation functions by a separation in \emph{real} space 
(i.e. non-distorted), generally denoted as $r$.  In practice, observations yield the distorted 
distance, and we refer to this redshift space separation as $s$.  When the angle subtended by 
$s$ is small (i.e. the plane-parallel approximation), we can decompose the redshift space distance 
into a line-of-sight ($\pi$) and projected separation ($r_p$) such that $ s = (\pi^2 + r_p^2)^{1/2} $.  
This effectively encapsulates the redshift distortion in the $\pi$ coordinate.

Redshift distortions \citep{jackson:72,sargent:77,peebles:80} produce two clearly observable effects 
on the galaxy distribution \citep[reviewed in][]{hamilton:98}.  At small scales, gravitational 
collapse becomes highly non-linear and large peculiar velocities produce elongated structures in the 
radial direction affectionately referred to as \emph{fingers-of-god} (the dominant effect at 
small $r_p$, see \citealt{jackson:72}).  Larger scales exhibit a flattening from linear infall, observed 
as less obvious structures perpendicular to the line-of-sight \citep{kaiser:87}.

We can parameterize the redshift space 2PCF in terms of the plane-parallel approximation, 
$\xi(s) \rightarrow  \xi(r_p,\pi)$, and integrate along $\pi$ to produce the projected 2PCF: 
\begin{equation} \label{eq:wp}
  w_p(r_p) = 2 \int_{0}^{\pimax} \xi(r_p, \pi) \mathrm{d} \pi \; .
\end{equation}
Physically, we project the 3D correlation function onto a 2D surface of the sky in a 
moving annulus of fixed width according to $\pimax$.
The projected 3PCF and its reduced form can be analogously defined as:
\begin{multline}   \label{eq:zeta_proj} 
  \zeta_{proj}(r_{p12}, r_{p23}, r_{p31}) = \\ 
    \iint \zeta(r_{p12}, r_{p23}, r_{p31}, \pi_{12}, \pi_{23}) \mathrm{d} \pi_{12} \mathrm{d} \pi_{23}  \; , 
\end{multline}
\begin{multline}   \label{eq:Qproj}
  Q_{proj}(r_{p12}, r_{p23}, r_{p31}) = \\
    \frac{ \zeta_{proj}(r_{p12}, r_{p23}, r_{p31}) }
	 { w_{p12} w_{p23} + w_{p23} w_{p31} + w_{p31} w_{p12} } \; .
\end{multline}
We extended our simplified notation such that $r_{p12}$ is the projected analog to 
$r_{12}$, and $w_{p12} = w_p(r_{p12})$.  We choose $\pimax = 20 \hmpc$ for the measurements 
we present in this work (discussed further in Appendix \ref{s:proj}).

Our notation is as follows. We show measurements of the 2PCF as a function of redshift space separation $s$. 
For projected space measurements, we keep the 2PCF characterized by a single dependent variable: $r_p$.  
The 3PCF is a function with three dependent variables and we continue the notation: $s$ or $r_p$ for 
redshift and projected space separations, respectively.  Finally, our use of $r$ is more general and can 
refer to separation in (theoretical) real, redshift or projected space.

\subsection{Estimating the Correlation Functions}
\label{ss:npcf}

We estimate the correlation functions using normalized pair counts for the 2PCF and triplets 
for the 3PCF.  We use a class of unbiased and minimal variance estimators \citep[see][]{szapudi:98} 
to optimally correct for edge effects.  The estimator we use for the 2PCF was first presented by 
\citet{landy:93}:
\begin{equation}\label{eq:est_ls}
  \widehat{\xi}_{LS} = \frac{DD - 2 DR + RR}{RR} \; ,
\end{equation}
where $DD$ denotes the normalized data-data pairs, $RR$ represents the normalized random-random pairs 
and $DR$ corresponds to the normalized cross count of data-random pairs. The $LS$ estimator was extended 
for all $n$-point correlation functions by \citet{szapudi:98}, and we estimate the 3PCF 
as
\begin{equation}\label{eq:est_ss}
  \widehat{\zeta}_{SS} = \frac{DDD - 3 DDR + 3 DRR - RRR }{RRR} \; .
\end{equation}
As before, $DDD$ represents the normalized count of data-data-data triplets, 
$DDR$ corresponds to data-data-random, etc.  

We construct random catalogs that are a factor of $5-10$ greater in density than the data.  
We find this ratio sufficient to keep the shot noise contribution of the random catalogs 
well below that of the data for all triplet counts, as well as small enough to be 
computational manageable.  We designed the random catalogs to have the exact redshift 
distribution as the data: for each galaxy, we generate $5-10$ points with random angular 
coordinates but identical redshifts. Finally, we match both the angular footprint and volume 
of the corresponding data samples.  

For our pair and triplet counts, we employ an exact \npoint\ calculator implemented in 
\ntropy, a parallel kd-tree framework \citep{gardner:07}.  The application, 
\texttt{ntropy-npoint}, utilizes an efficient algorithm developed by 
\citet{moore:01} and discussed in \citet{gray:04}.  Their original implementation, 
\texttt{npt}, is publicly available and has been used to investigate the 3PCF 
\citep{nichol:06,kulkarni:07,marin:08}.  The independent \npoint\ implementation based on 
\ntropy\ shows enhanced runtime performance, true parallel capability, and the ability to 
search directly in projected coordinates of $r_p,\pi$ \citep[see details in][]{gardner:07}.
We define the coordinates $\pi$ and $r_p$ for each point pair, where the 3D separation 
is $\vec{r}_{12} = \vec{x}_1 - \vec{x}_2$. We use the unit direction to the midpoint of 
the separation, $\widehat{r}_{m}$ to calculate $\pi = \widehat{r}_m \cdot \vec{r}_{12}$ and 
find $r_p^2 = r^2_{12} - \pi^2$.

We stress that the ability to utilize massively parallel computing platforms proved extremely 
important, as the projected 3PCF required almost two orders of magnitude more time to 
compute than the spatial 3PCF.  We utilized hundreds of processors at a time and required 
$\sim\!300,000$ CPU hours to complete the $n$-point calculations presented here.

\subsection{Triplet Configurations and Binning Scheme}
\label{ss:config}

The full 3PCF is a function of three variables that characterize both the size and shape 
of triplets.  A natural and unique description of a triplet is the length of each 
side of the triangle that connects the points: $r_{1}$, $r_{2}$ and $r_{3}$ where 
connectivity is assumed by $\vec{r}_{3} = \vec{r}_{1} + \vec{r}_{2}$.  


In perturbation theory, it is most common to see triangles described by two side lengths 
($r_{1}$ and $r_{2}$) and their opening angle ($\theta$) defined by the cosine rule: 
\begin{equation}\label{eq:cos_rule}
	\cos \theta = \frac { r_{1}^2 + r_{2}^2 - r_{3}^2 }{ 2 r_{1} r_{2} } \; .
\end{equation} 
We find this last characterization the most natural and intuitive and will use it to describe our 
3PCF measurements.  When $\theta \approx 0$ or $\theta \approx \pi$, we refer to triplets as ``collapsed'' or 
``elongated'' respectively, both of which have two sides being very close to co-linear.  As $\theta$ approaches 
$\pi/2$ the triplet forms a right triangle which we will refer to as a ``perpendicular'' configuration.  
The triangle shape, or \emph{configuration} dependence, describes a function of $\theta$ where 
``\emph{strong} configuration dependence'' means a significant amplitude difference in the reduced 
3PCF, $Q(r_1,r_2,\theta)$, between co-linear and perpendicular configurations and 
``\emph{weak} configuration dependence'' shows little or no change of in $Q(r_1,r_2,\theta)$ with $\theta$.  
This terminology relates to that used by \cite{GS05}. 

While it might appear this description is both trivial and pedantic, we caution that these mappings 
are non-linear. Triangle descriptions such as side-side-side and side-side-angle remain completely 
equivalent for exact values when the triangle vertices are infinitesimal points.  However, these transformations 
will not be exact for small volumes, as in the case of estimating correlation functions where we accumulate 
counts within bins of a finite width. Care must be taken to prevent significant discrepancies in 3PCF 
measurements if the parameterization used for the calculation differs from the one that is 
modeled.  We carefully consider this subtlety to correctly interpret our measurements. 

We choose ($r_1$, $r_2$, $\theta$) to parametrize our 3PCF measurements, although we measure the
3PCF using bins defined by ($r_1, r_2, r_3$) and use \eqref{eq:cos_rule} to convert $r_3$ to $\theta$.
Even though we treat the reduced 3PCF as a function of three variables, $Q(r_1, r_2, \theta)$, we may
often denote it as $Q(\theta)$ of even $Q$ for simplicity.

A good choice of binning is not straightforward for the configuration dependence of the 3PCF.
Since neighboring bins need to be tightly packed to measure $Q(\theta)$, choosing a bin-width based 
on $\log r$ will be too large and insufficient to resolve a slow variation.  On the other hand, a small 
bin-size will quickly become under-sampled, as triplets are characterized by three variables.  We choose 
to employ linearly spaced bins in $\theta$, and use the cosine rule in \eqref{eq:cos_rule} to define the 
midpoint $r_{3}$.  We choose a bin-width as a fraction, $f$, of the measured scale, $r$, such that 
$\Delta_r = f \times r$ and a bin at $r$ is measured between 
$(r - \frac{\Delta_r}{2}, r + \frac{\Delta_r}{2})$.  
This scheme was independently found to be useful in a theoretical study of the 3PCF by \citet{marin:08}.
We compare measurements between three volume-limited galaxy samples which have very different number 
densities (see Table~\ref{t:gal_samples}), and select a single fraction $f$ to apply to all samples 
and scales for a fair comparison.  We set the value of the fractional bin-width by requiring a 
``reasonable sampling'' of the sparsest dataset: the BRIGHT galaxy sample. 
Using a fractional bin-width of 25\% ($f = 0.25$) results in several hundred triplet counts at the 
smallest scales ($r_1 = 3 \hmpc$), and we use this $f$ value for all measurements.  This produces 
relatively wide bins with some that physically overlap (the same triplet is counted more than once). 
We note two observable consequences: 
(1) a slight damping of the 3PCF near $\theta \approx \pi$, and 
(2) an induced correlation between some neighboring $\theta$ bins.  
Qualitatively, this should not pose a problem since all comparisons use the exact same scheme 
(including \nbody\ results).  Quantitatively, we account for this by using the full covariance matrix.  
We further discuss the implications of binning in the appendix, \S~\ref{ss:binning}.

\subsection{Estimating the Covariance Matrix}
\label{ss:covar_est}

We measure the correlation between measurements by empirically calculating the covariance matrix.  
Given a number of realizations, $N$, a residual on $Q$ can be defined as 
\begin{equation}\label{eq:del}
  \Delta_i^k = \frac{ Q_i^k - \bar{Q}_i }{ \sigma_i } \; ,
\end{equation}
for each realization ($k$) and bin ($i$) given a mean value ($\bar{Q}_i$) and variance ($\sigma_i^2$) 
for each bin over all realizations.  We use $Q$ as a general placeholder for any measured statistic
(2PCF, 3PCF, etc).  A covariance matrix can be estimated from the data itself using a 
\emph{leave one out cross validation} method, more commonly referred to as jackknife 
re-sampling \citep[see further description in][]{lupton:01}.

The number of jackknife samples (which we denote with $N$) are not independent realizations, and 
the jackknife covariance matrix can be estimated by:
\begin{equation}\label{eq:cov_jack}
  \mathcal{C}^{(jack)}_{ij} = \frac{(N - 1)^2}{N} \mathcal{C}_{ij} = \frac{N - 1}{N} \sum_{k=1}^{N}\Delta_i^k \Delta_j^k \; , 
\end{equation}
where $\mathcal{C}_{ij}$ denotes the typical unbiased estimator of the covariance when 
computed on jackknife samples.  Jackknife re-sampling has been shown to be reliable on scales up to 
$30 \hmpc$ for the 2PCF on spectroscopic galaxy samples when compared with independent mock catalogs 
\citep{zehavi:02, zehavi:04, zehavi:05}.  However, a more elaborate study by 
\citet{norberg:09} highlights potential problems and caveats of using such an ``internal'' 
estimate of the errors.  We defer questioning the validity of jackknife re-sampling methods for 
estimating the covariance of the reduced 3PCF to a companion paper (McBride et al. 2010). 
They compare jackknife estimates to those from mock galaxy catalogs generated from \nbody\ 
simulations. They conclude that while jackknife error estimates do not exactly agree with mocks, 
they appear sufficient for the analysis presented here (McBride et al 2010). 

We generate our jackknife samples using the pixelization scheme of STOMP 
\footnote{http://code.google.com/p/astro-stomp/}. This code can account for the irregular 
geometry of the SDSS data, and has been used to quantify errors in angular clustering 
\citep{scranton:05}.
The jackknife regions are selected to maintain equal unmasked area on the sky.  In the 
SDSS samples we use, survey depth does not vary over the sky, which makes equal area 
consistent with equal volume.  We want to resolve the covariance of $Q(\theta)$ for $15$ 
bins in $\theta$ between $0$ and $\pi$, therefore using less than $15$ jackknife samples 
should formally result in a singular covariance matrix.  We choose $30$ jackknife samples 
where the regions are each $\sim180$ square degrees in unmasked area (see discussion in 
McBride et al. 2010 where they investigate varying the number of jackknife samples for 
these $Q(\theta)$ measurements).

\section{SDSS Measurements}
\label{s:results}

\subsection{Two-Point Correlation Function}
\label{ss:2pcf}

\begin{figure*}[ht]
  \centerline{
    \includegraphics[angle=0,width=0.4\textwidth]{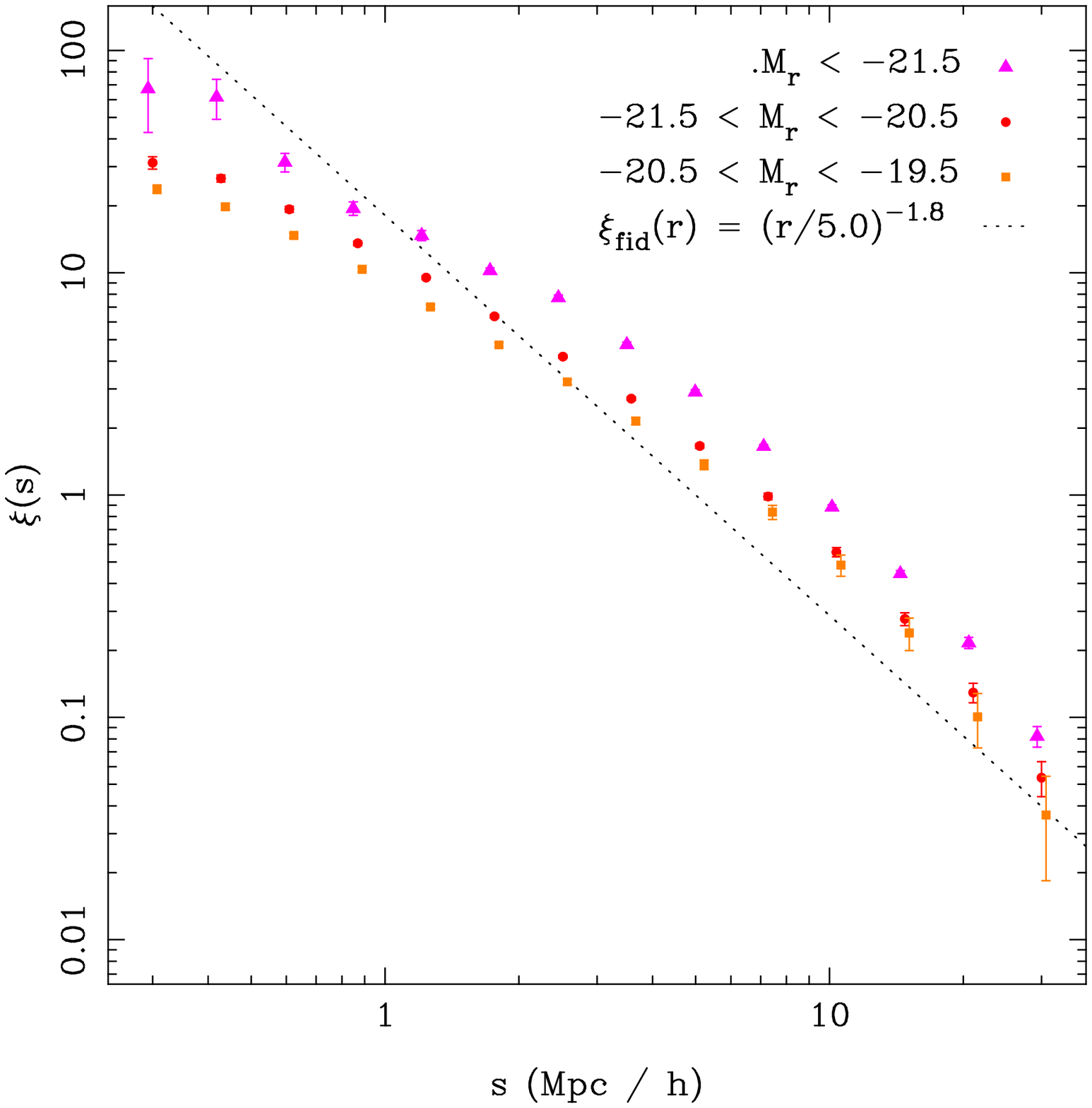}
    \includegraphics[angle=0,width=0.4\textwidth]{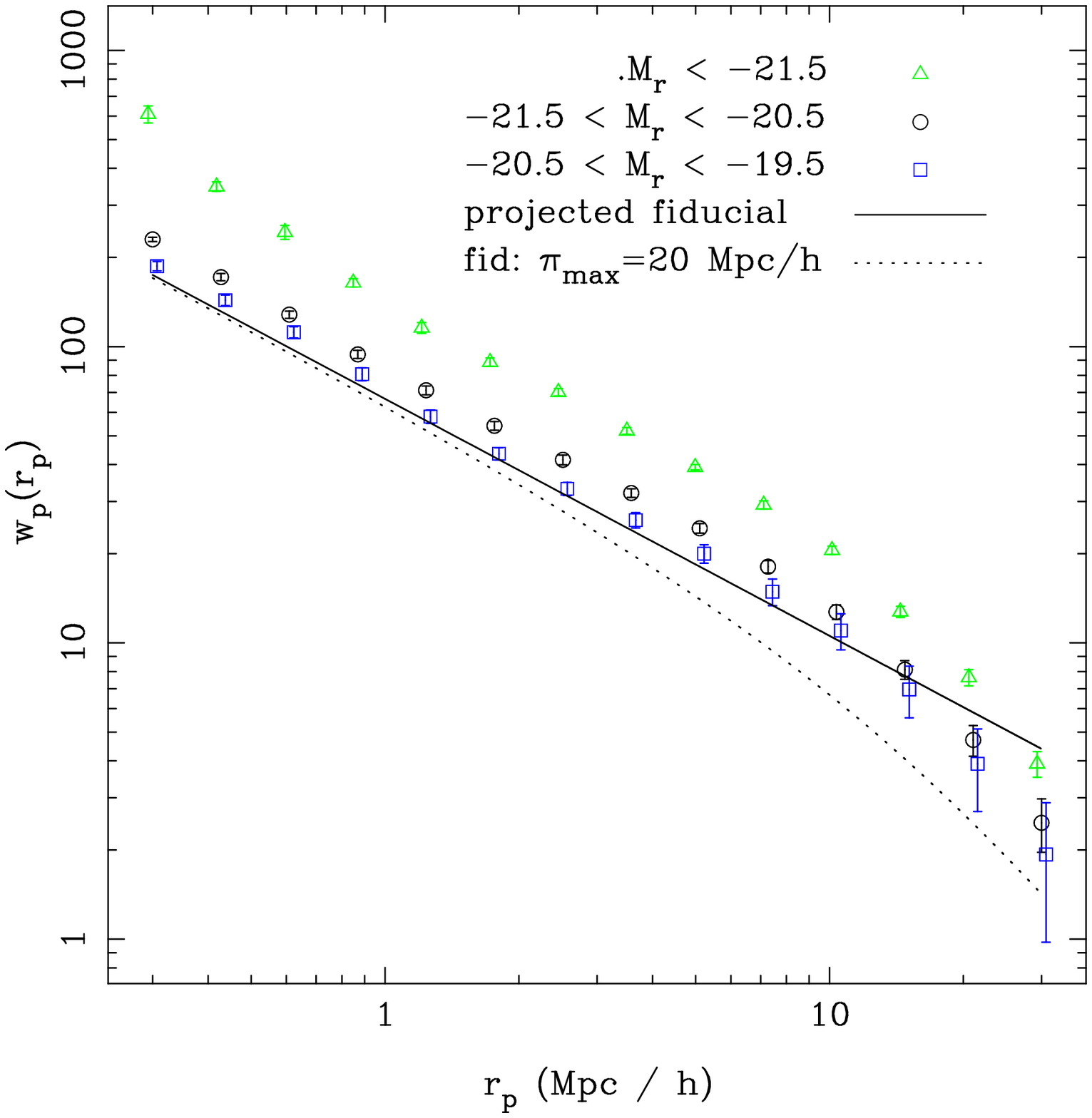}
  }
  \caption[2PCF of SDSS samples]{ 
    The redshift space 2PCF, $\xi(s)$ in the left panel, and projected 2PCF, $w_p(r_p)$ in 
    the right panel, for our three SDSS samples. Brighter galaxy samples exhibit more 
    clustering (e.g. higher amplitude) than fainter samples, in agreement with published 
    analyses of SDSS galaxies \citep{zehavi:05}.  
    Error bars denote $1$$\sigma$ uncertainties calculated from 30 jackknife samples.
  } \label{f:2pcf}
\end{figure*}

We first investigate the 2PCF of our samples.  While we later focus our clustering analysis on
the reduced 3PCF, $Q(r_1,r_2,\theta)$, it is instructive to directly examine the 2PCF since $Q$ is
the ratio of the connected 3PCF, $\zeta(r_1,r_2,\theta)$ divided by products of the 2PCF, $\xi(r)$, 
such that $Q \propto \zeta / \xi^2$ as shown in \eqref{eq:Q}.  We choose equal width bins in 
$\log r$ to measure the 2PCF.

The galaxy correlation function was first shown to be a power law by \citet{totsuji:69} using 
galaxy data from the Lick Observatory (Shane and Wirtanen, 1967).
Along with our measurements, we plot a \emph{fiducial} power law model, 
\begin{equation} \label{eq:xi_pl}
  \xi(r) = \left(\frac{r}{r_o}\right)^{-\gamma} \; ; \; \xi_{fid} = \left(\frac{ r }{5 \hmpc}\right)^{-1.8} , 
\end{equation}
which we adopt for comparison from a detailed study of clustering in SDSS galaxies by
\citet{zehavi:05}.  Given the precision of modern measurements, recent work has noted significant
departures of the galaxy 2PCF from a power law \citep{zehavi:04} where the data are better described
by the halo model \citep[reviewed in ][]{cooray:02}.  Nevertheless, a power law provides a simple
and convenient comparison.

We plot the 2PCF in both redshift space, $\xi(s)$, and projected space, $w_p(r_p)$, for 
our three galaxy samples in Figure~\ref{f:2pcf}.  At larger scales, $\xi(s)$ runs almost 
parallel to the \emph{fiducial} model for all galaxy samples.  
We note the reduction in strength of $\xi(s)$ at separations below $\approx 5\hmpc$ in 
Figure~\ref{f:2pcf}.  At these scales this is an effect of redshift distortions where 
the fingers-of-god make galaxy pairs at close separations less likely and reduce the 
small scale correlation function. 

In contrast to $\xi(s)$, the $w_p(r_p)$ does not exhibit the same reduction in power, and 
remains close to a power law for small projected separations.  This confirms our 
expectations that the projected $w_p(r_p)$ should be less affected by redshift 
distortions -- especially at the smaller projected scales (see Figure~\ref{f:2pcf}).  At 
large $r_p$, $w_p(r_p)$ shows a reduction of power due to truncating the $\pi$ integration 
at $\pimax = 20\hmpc$.  We show this effect by using our fiducial model.  Assuming a perfect 
power law in real space, we depict the resulting $w_p(r_p)$ with two lines in Figure~\ref{f:2pcf}, 
where the dotted line uses the same $\pimax$ as the data and solid line depicts the full 
projection (i.e. $\pimax = \infty$).  

In both $\xi(s)$ and $w_p(r_p)$ we note that brighter galaxies exhibit stronger 
clustering when compared to fainter galaxies.  We can understand this in terms of a simple 
galaxy-mass bias parameter \citep{fry:93}, where $\xi \propto b^2$ and the linear bias, $b$, increases 
with galaxy luminosity; this has been studied in detail for SDSS galaxies\citep{zehavi:05, zehavi:10}.  
Our measurements are in agreement with their analysis.  The galaxy 2PCF depicts a higher amplitude 
than the fiducial model, which is consistent with the luminosity of our galaxy samples.

\subsection{Equilateral Three-Point Correlation Function}
\label{ss:Qeq}

\begin{figure*}
  \centerline{
    \includegraphics[angle=0,width=0.4\textwidth]{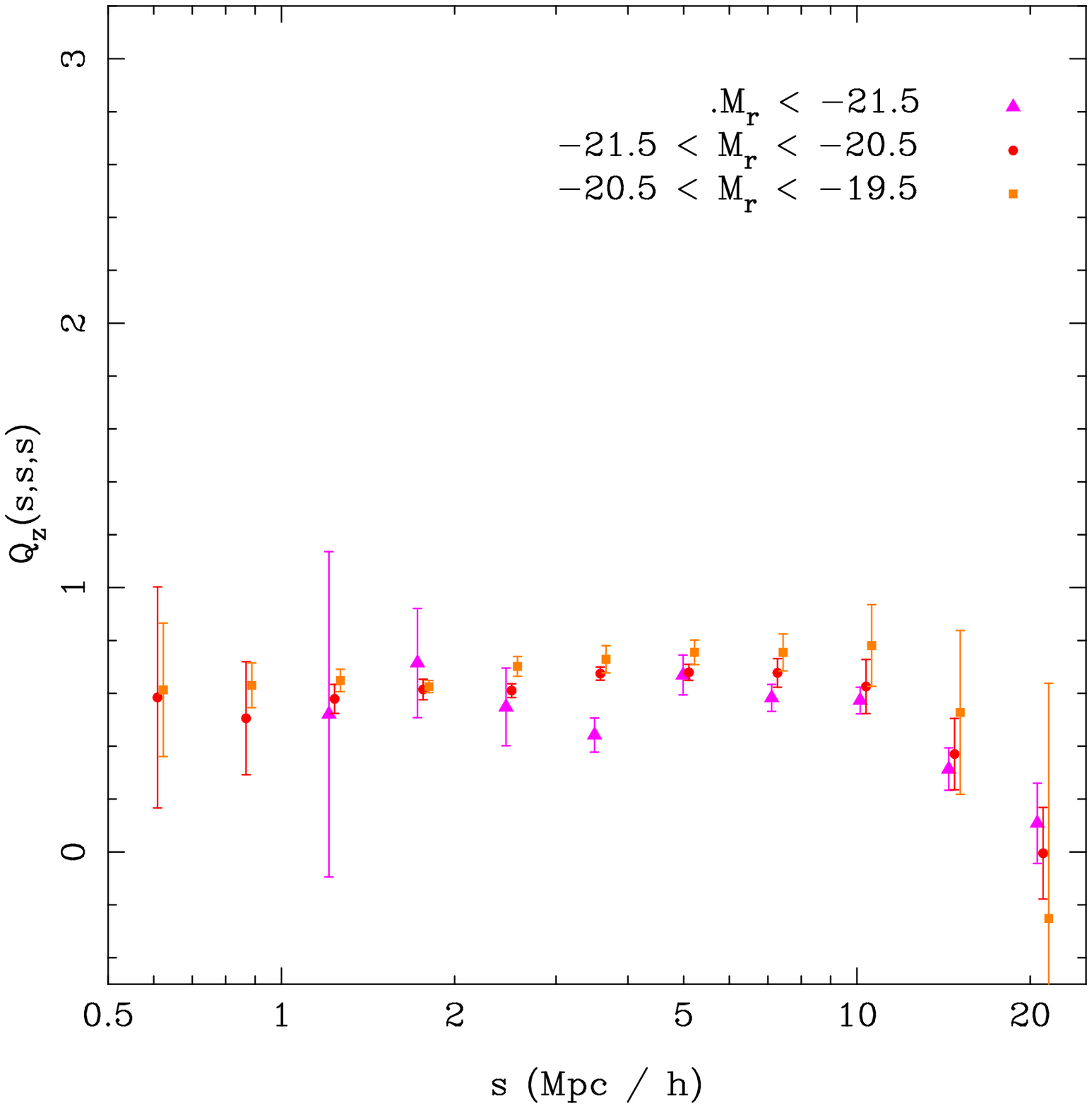}
    \includegraphics[angle=0,width=0.4\textwidth]{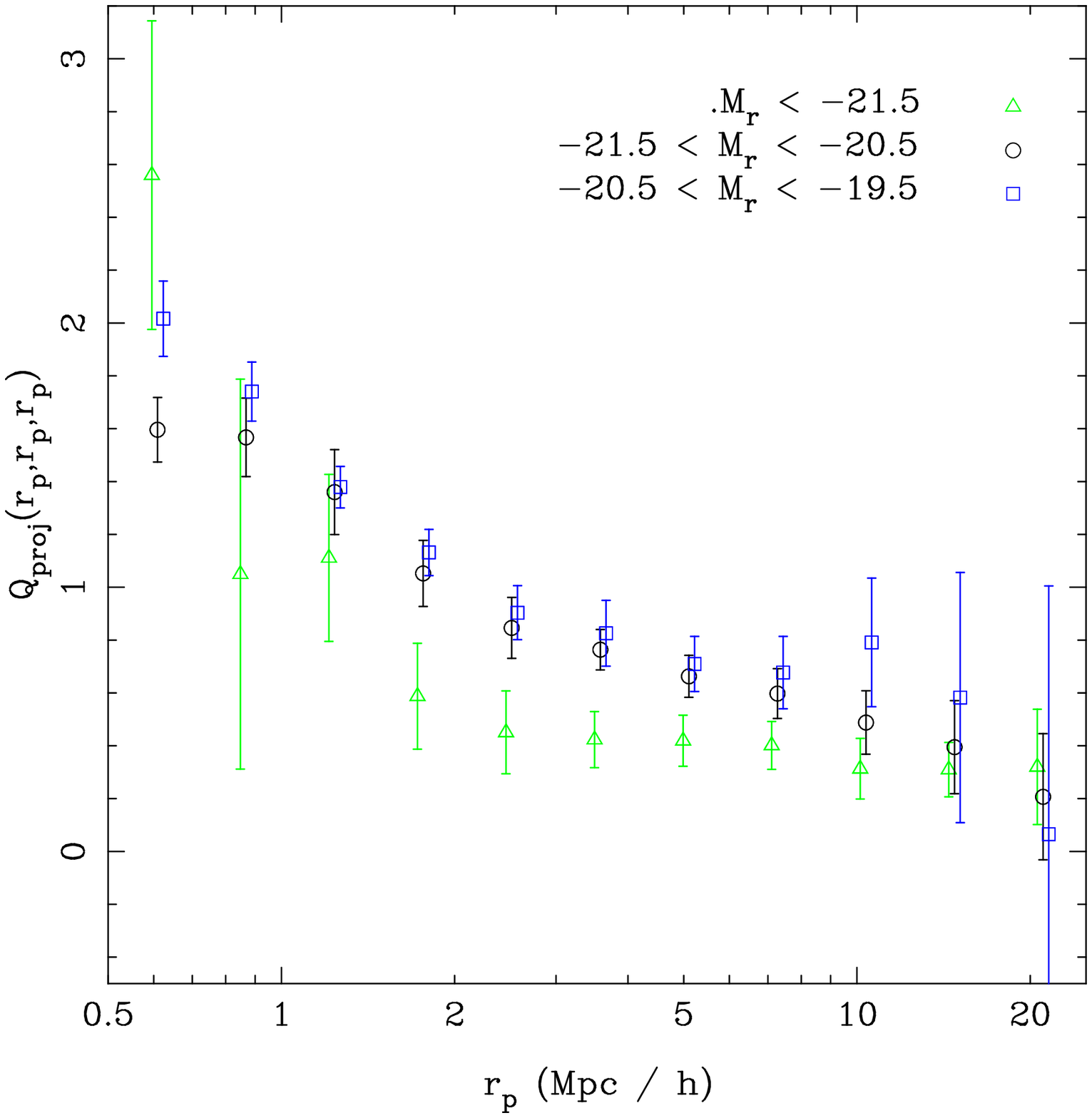}
  }
  \caption[Equilateral 3PCF for SDSS DR6 galaxies]{ 
    The equilateral reduced 3PCF, $Q_{eq}(r)$, for our three SDSS galaxy samples, both in redshift space 
    (left) and projected space (right).  The equilateral reduced 3PCF is characterized by a single scale 
    for simplicity: $Q_{eq}(r) = Q(r,r,r)$. We choose a bin-size to be equal width in $\log r$ analogous 
    to the binning we used for the 2PCF. 
    Error bars denote $1$$\sigma$ uncertainties calculated from 30 jackknife samples.
  } \label{f:Qeq}
\end{figure*}

The simplest analog to the 2PCF is the equilateral reduced 3PCF, $Q_{eq}(r)$.
As each side of the triangle formed by a triplet corresponds to the same scale 
($r_1 = r_2 = r_3 = r$), $Q_{eq}$ can easily be characterized by a single 
separation:
\begin{equation}\label{eq:Qeq}
  Q_{eq}(r) = \frac{\zeta(r,r,r)}{3 \; \xi(r)^2} \; .
\end{equation}
$Q_{eq}$ is not sensitive to shape, but it provides a clear sense of scale dependence 
in the reduced 3PCF.  When $Q_{eq}\!\approx 1$, the number of triplets exactly correspond to 
those expected from the 2PCF; when $Q_{eq}$ is above or below $1$, there are more or less 
triplets, respectively. Like the 2PCF, we choose bins of equal width in $\log r$.  

We show $Q_{eq}$ in both redshift and projected space for our three galaxy samples
in Figure~\ref{f:Qeq}.  In redshift space, $Q_{eq}$ appears flat and never exceeds $1$,
showing very little difference in terms of luminosity. In the projected measurement, small
scales reveal $Q_{eq} > 1$ until $r_p \approx 2-3\hmpc$, where it approximately reproduces
the amplitude of the redshift space measurement.  We expect larger scales to dip below
one, as $\zeta(r)$ drops more rapidly than pairs in the 2PCF for the same scale.  
Intuitively, the number of equilateral triplets of a set side length become more 
rare than the number of available pairs of the same separation length.

On small scales redshift distortions destroy small scale triplets in a manner similar to what we
discussed for pairs in the 2PCF.  We notice that redshift space $Q_{eq}$ remains flat in
Figure~\ref{f:Qeq}, but the projected $Q_{eq}$ recovers power at small $r_p$.  This suggests the
flattening of the redshift space measurement is a result of redshift distortions. This
interpretation is supported by a theoretical comparison of the real and redshift space $Q_{eq}$ in
\citet{marin:08} (see their Figure~1). We notice a slight variation of the measurement with sample
luminosity, but not nearly as pronounced as seen in the 2PCF, and only significant for a few points.
At the largest scales ($r \ge 10\hmpc$), the reduction in amplitude of $Q_{eq}$ and increased size
of the errors suggest that we are limited in signal by the finite volume of the galaxy sample. The
effects appear worst for the faintest volume-limited sample, which is expected since it also occupies
the smallest volume in comparison with the larger volume brighter samples (see
Table~\ref{t:gal_samples}). Finite volume effects have been shown to cause a rapid reduction in
amplitude for higher order measurements of clustering \citep{scoccimarro:00}.

\subsection{Configuration Dependence of reduced 3PCF}
\label{ss:Q_conf}

We focus our investigation on the shape or \emph{configuration} dependence of the reduced 
3PCF, $Q(r_1,r_2,\theta)$.  The reduced 3PCF was analyzed in redshift space by 
\citet{kayo:04} and \citet{nichol:06} using a previous release of SDSS galaxy data: 
``sample12'' -- a version between the first and second data releases covering $\sim\!2406$ 
square degrees.  
Unlike these previous studies, we restrict our analysis only to volume-limited galaxy 
samples, and jointly investigate redshift and projected space measurements. 
We measure the reduced 3PCF for SDSS DR6 galaxies at three scales, which we specify by the 
smallest side of the triangle, $r_1$.  We choose $r_2$ so that the ratio of the first two 
sides stays fixed such that $r_2 / r_1 = 2$.  We then measure the reduced 3PCF as a function 
of opening angle between these two sides, $\theta$, regularly spaced between $0$ and $\pi$.  
The resulting scale we probe varies from $r_1$, when $\theta = 0$ (i.e. ``collapsed'') to 
$(r_1 + r_2)$, when $\theta = \pi$ (i.e. ``elongated'').  Specifically we measure triplets 
on scales between $3-9$, $6-18$ and $9-27 \hmpc$ corresponding respectively to $r_1 = 3,6$ 
and $9 \hmpc$.

\begin{figure*}
  \centerline{
    \includegraphics[angle=270,width=\linewidth]{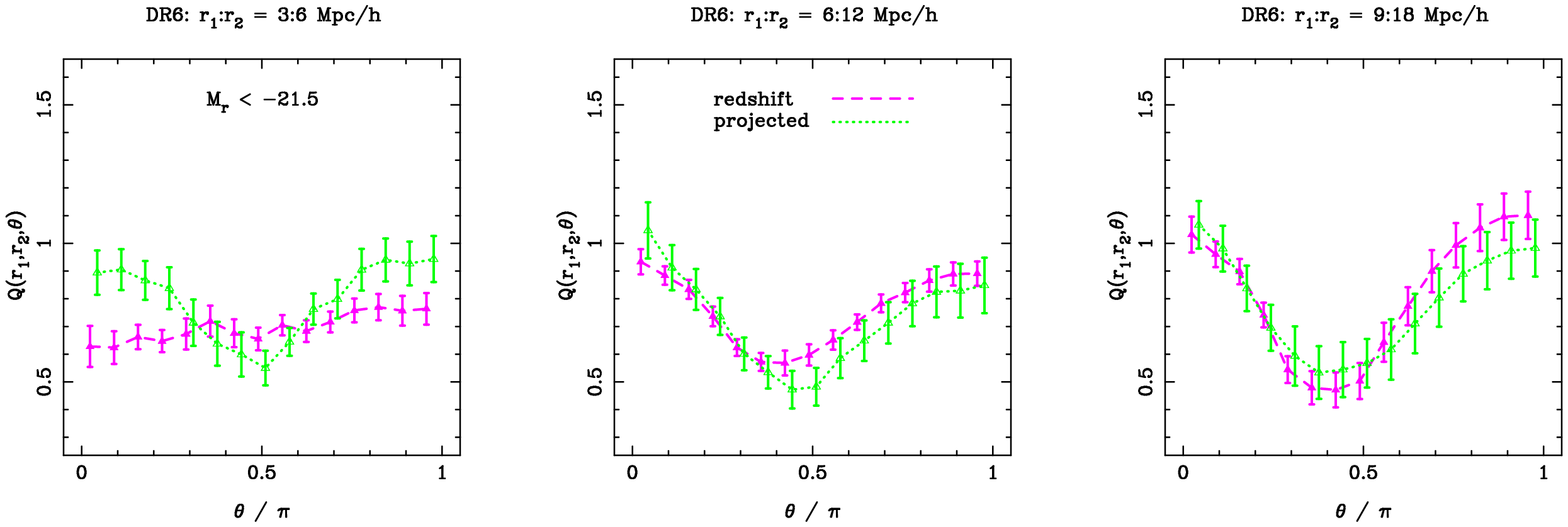}
  } \centerline{
    \includegraphics[angle=270,width=\textwidth]{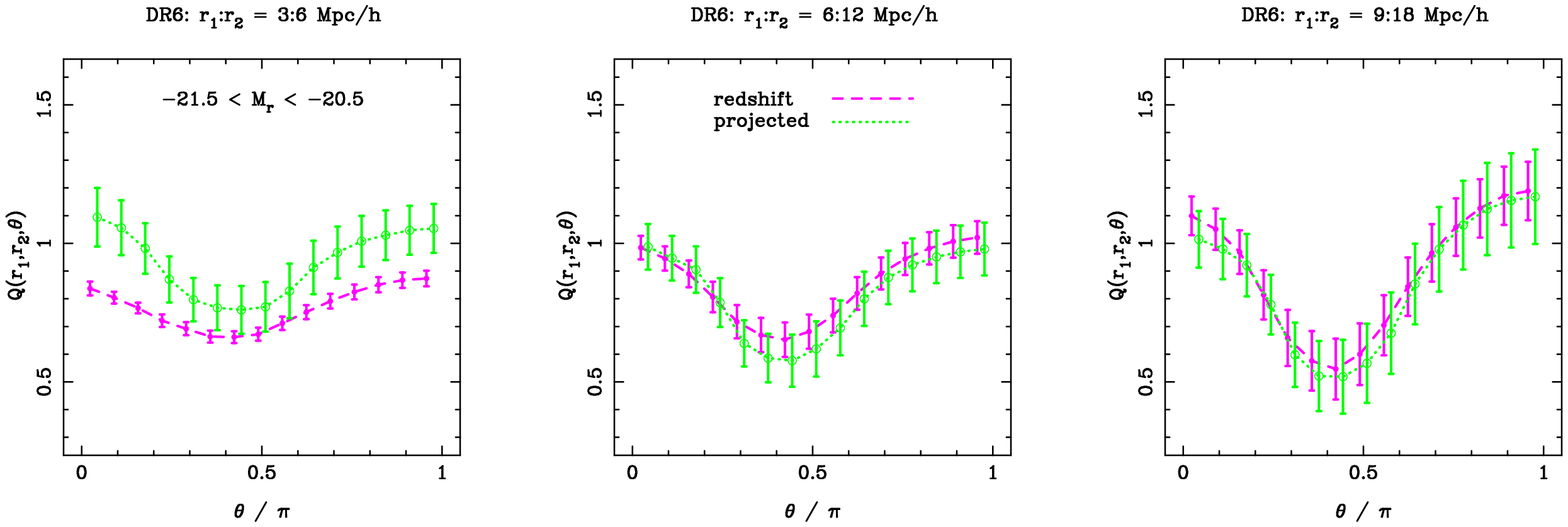}
  } \centerline{
    \includegraphics[angle=270,width=\textwidth]{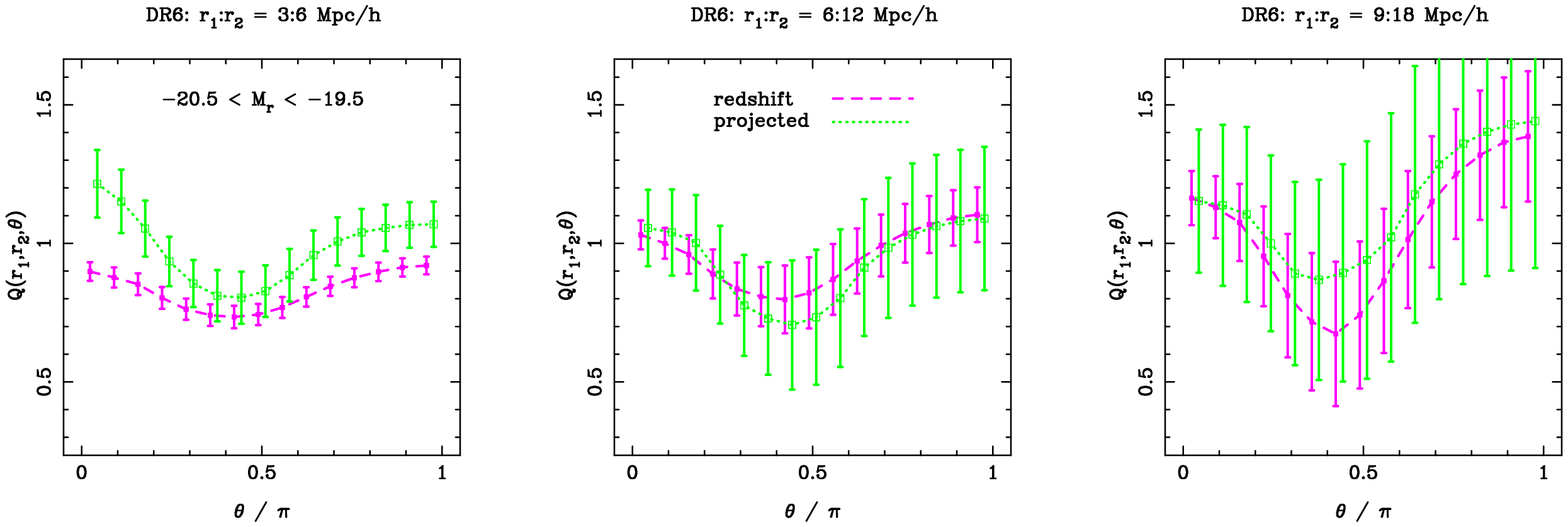}
  }
  \caption[Comparison of redshift and projected space reduced 3PCF]{
    The reduced 3PCF on SDSS DR6 galaxies showing both redshift space and projected 
    measurements.  The filled symbols are redshift measurements, and the hollow symbols 
    are projected. Triangles correspond to $M_r < -21.5$, circles with 
    $ -21.5 < M_r < -20.5 $ and squares with $ -20.5 < M_r < -19.5 $.  
    The three columns are different scales, specified by the first side of the triangle 
    ($r_1$) representing the smallest scale measured. 
    Error bars denote $1$$\sigma$ uncertainties calculated from 30 jackknife samples.
  } \label{f:Q_typecmp}
\end{figure*}

We introduce the configuration dependence for each sample and scale in Figure~\ref{f:Q_typecmp}, 
where we overlay the reduced 3PCF in both redshift and projected space.  On smaller scales (left 
side of Figure~\ref{f:Q_typecmp}), we see weaker configuration dependence in the reduced 
3PCF (i.e. smaller change in $Q(\theta)$ with $\theta$) than at the larger scales (right side).  

We notice a ``V-shape'' at $r_1 = 9 \; \hmpc$, present in both redshift and projected
space measurements of Figure~\ref{f:Q_typecmp}.  We interpret this as a statistical
signal of filamentary structure that becomes more pronounced as the scale increases: 
an over-abundance ($Q > 1$) of co-linear triangles with an under-abundance 
($Q < 1$) of perpendicular configurations.  This characteristic V-shape is predicted 
from gravitational perturbation theory \citep{lss_review} as well as results from 
simulations \citep{GS05,marin:08}.  

We identify the effect of redshift distortions on the small scales in the redshift space 
reduced 3PCF in two ways: (1) almost no configuration dependence in $Q_z(\theta)$ and 
(2) $Q_z(\theta) < 1$ for all configurations, showing a deficiency of triplets due to 
non-linear velocities stretching compact structures beyond the scales probed (i.e. fingers-of-god).  
On small scales, we notice that $Q_{proj}$ has both a higher amplitude and exhibits a 
stronger configuration dependence in comparison with $Q_z$.  When comparing with the 2PCF 
(Figure~\ref{f:2pcf}), the projected measurements are larger in amplitude.  However, 
$Q_{proj}$ is normalized by $w_p(r_p)$, so we expect $Q_{proj}$ to match $Q_z$ and 
remain close to $1$ in value.  The increased amplitude of $Q_{proj}$ is attributed to 
recovering signal that was lost to $Q_z$ due to redshift distortions.  We consider the 
stronger configuration dependence of $Q_{proj}$ reflective of the true galaxy distribution, 
as a simple projection should not increase the configuration dependence.  We conclude that 
the projected measurement recovers signal that is destroyed in the redshift space measurement 
due to non-linear distortions.  However, we expect that our choice of $\pi_{max} = 20 \hmpc$ 
will permit redshift distortions to affect these projected measurements, albeit in a 
reduced capacity with respect to redshift measurements.  As the scale increases beyond 
$9 \hmpc$, $Q_{proj}$ and $Q_z$ generally converge (right column of Figure~\ref{f:Q_typecmp}).  

We see a dramatic difference in the size of uncertainties between these measurements.  
For each individual sample and statistic, there are two important aspects to consider.  
First, the small scales become sensitive to sampling where a galaxy sample with a lower 
number density will be less well determined (shot noise limited).  Second, the larger 
scale measurements will show more sensitivity to volume effects (being dominated by sample 
or cosmic variance).  We see this same behavior for $Q_{eq}$ in Figure~\ref{f:Qeq}.

The uncertainties increase with larger scales, especially for the fainter galaxy samples.  
We actually expect the growth of errors from the ``ratio'' static of the reduced 3PCF since 
we predict the reduced 3PCF to be approximately unity at all scales.  We interpret the general 
increase of errors on $Q$ as the higher order 3PCF (numerator in $Q$) becoming uncertain at 
a much faster rate than the 2PCF (denominator).  

We notice the uncertainties on the FAINT sample appear much larger than measurements on the 
two brighter samples, especially at larger scales.  This is not Poisson uncertainty as the 
sample contains over 76,000 galaxies, a factor of two greater than the BRIGHT sample.
The FAINT galaxy sample occupies a much smaller volume, and a few large structures 
dominate the errors in the large scale measurements (see discussion in \S\ref{ss:sstruct}).  
The observed increase of uncertainties with fainter galaxy samples is generally a 
consequence of the smaller volume being probed by the fainter volume-limited samples 
(see Table~\ref{t:gal_samples}).  

The projected $Q(\theta)$ shows larger uncertainties than redshift space measurements 
at all scales.  This is a natural side effect of the projection: a given scale in $r_p$ 
represents a lower bound on the redshift space scale, $s$, where the upper 
bound is $ s = (\pimax^2 + r_p^2)^{1/2}$.  Since $Q(\theta)$ at any given value 
of $r_p$ is sensitive to scales greater than the same value of $s$, it will encode 
properties of the larger scales.  As the uncertainty grows with scale, $Q_{proj}$ will 
reflect higher uncertainties than $Q_z$ when $r_p \approx s$.

\subsubsection{Luminosity Dependence}
\label{ss:Q_lum}

\begin{figure*}
  \includegraphics[angle=270,width=\textwidth]{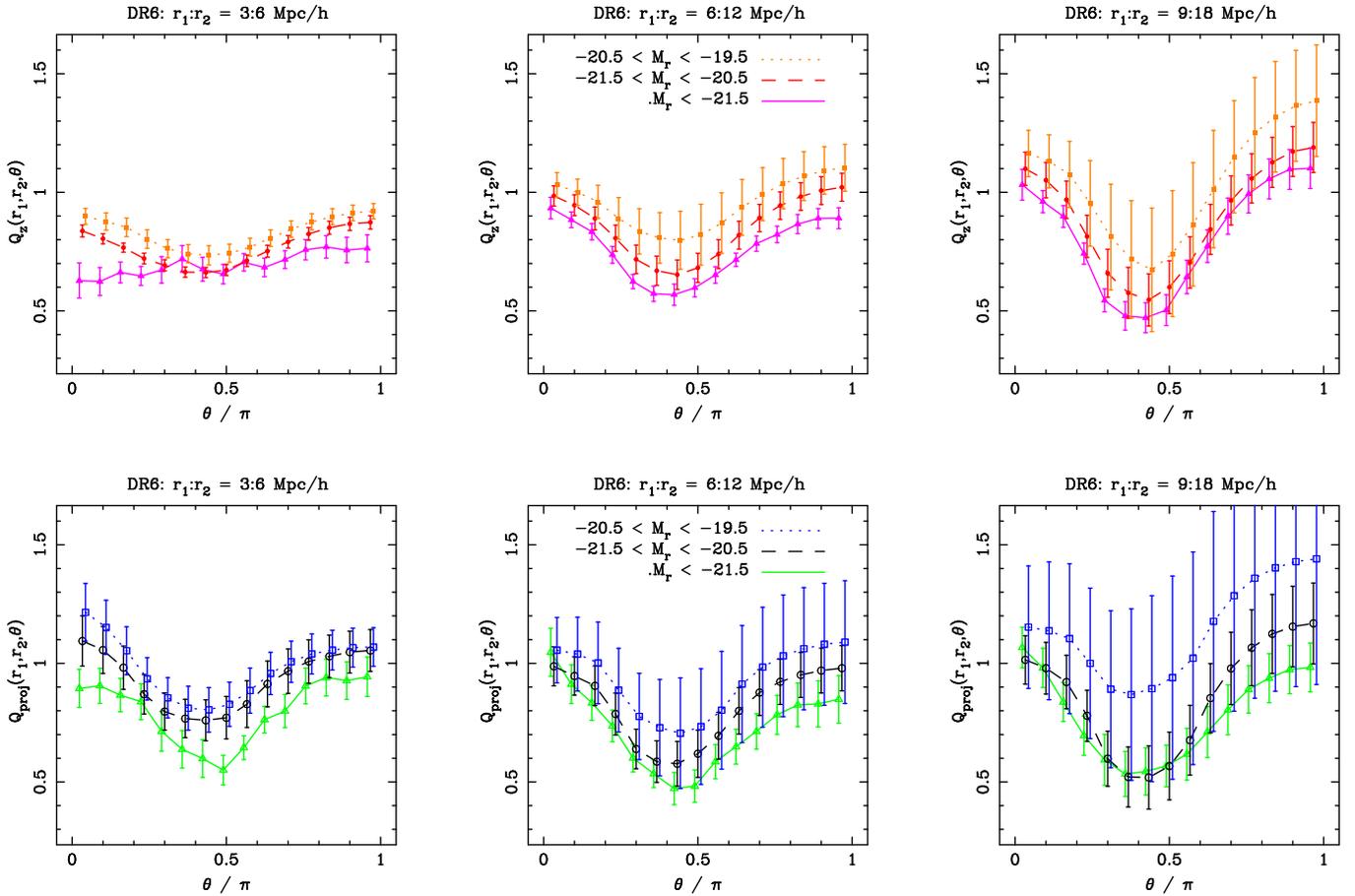}
  \caption[Luminosity dependence of reduced 3PCF from SDSS DR6 galaxies]{
    The reduced 3PCF on SDSS DR6 galaxies, comparing the three samples of different 
    magnitude limits.  Triangles correspond to $M_r < -21.5$, circles with 
    $-21.5 < M_r < -20.5 $, and squares with $ -20.5 < M_r < -19.5 $.  The top row 
    contains redshift space measurements, and the bottom row depicts the projected 
    measurements.  The three columns are different scales, specified by the first side of 
    the triangle ($r_1$) representing the smallest scale measured. 
    Error bars denote $1$$\sigma$ uncertainties calculated from 30 jackknife samples.
  } \label{f:Q_lumcmp}
\end{figure*}

To highlight the effect of luminosity on the reduced 3PCF, we present the measurements of 
all three samples in Figure~\ref{f:Q_lumcmp}.  As we noted in the 2PCF, the luminosity of 
the sample affects the clustering strength. However, we see the opposite behavior in 
$Q(\theta)$ as brighter galaxies show \emph{lower} values of $Q$ than fainter galaxies.  
Also remember that in the 2PCF (Figure~\ref{f:2pcf}) brighter galaxies showed a stronger 
segregation with galaxy luminosity.  If brighter galaxies exhibit stronger clustering, 
they will have a correspondingly higher linear bias $b$.  Since $\xi \propto b^2$, the 
amplitude will be more sensitive to changes in $b$ as opposed to the reduced 3PCF, where 
$Q \propto 1 / b$.  The difference of $Q(\theta)$ with sample luminosity appears constant, 
or slightly increases, with scale. 

Physically, we expect bright galaxies to predominantly live in galaxy groups or clusters
and centered at the ``knots'' in LSS. Fainter galaxies will be more populous in the field 
and overdense filaments, which might show more configuration dependence in the reduced 3PCF.  
The uncertainties on these measurements prevent a clear demonstration of this effect -- 
though we might speculate that a few measurements at $r_1 = 9 \hmpc$ hint at fainter 
samples showing a stronger configuration dependence (e.g. LSTAR vs BRIGHT in lower right 
panel of Figure~\ref{f:Q_lumcmp})


\subsubsection{Color Dependence}
\label{ss:Q_color}

We investigate the color dependence of the reduced 3PCF using two volume-limited samples, 
designed to have unit magnitude bins with limiting $r$-band magnitudes below and above
$\Lstar$.  We divided the galaxies into ``red'' and ``blue'' sub-samples to probe 
the color dependence as described in \S\ref{s:data}.  

\begin{figure*}
  \includegraphics[angle=270,width=\textwidth]{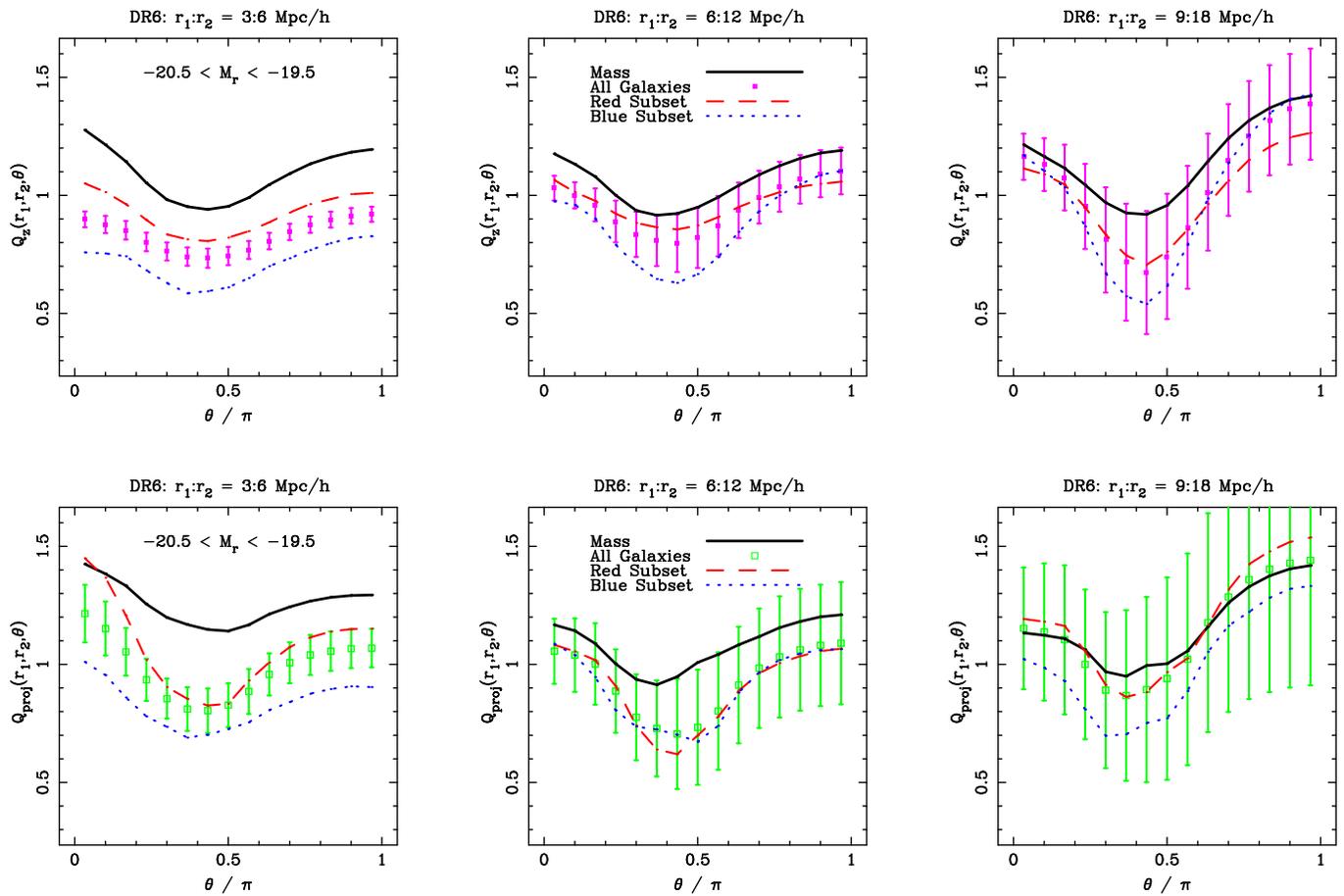}
  \caption[Reduced 3PCF of the $-20.5 < M_r < -19.5$ DR6 galaxy sample with color separation]{
    We show the configuration dependence of the reduced 3PCF for our FAINT galaxy sample with 
    $ -20.5 < M_r < -19.5 $.  The sample is divided into ``red'' (red, dashed line) and 
    ``blue'' (blue, dotted line) sub-samples based on SDSS $g-r$ color (see sample description for details).  
    The top row consists of redshift space measurements, and the bottom row depicts projected measurements.  
    The three columns correspond to different scales specified by the first side of the 
    triangle ($r_1$) representing the smallest scale measured.
    The black line denotes measurements on dark matter in the Hubble Volume simulation, but matching the 
    selection of the galaxies and includes redshift distortions.
    Error bars denote $1$$\sigma$ uncertainties calculated from $30$ jackknife samples.
  }\label{f:Q_mb19_5}
\end{figure*}

We examine the reduced 3PCF of the FAINT sample ($-20.5 < M_r < -19.5$) in 
Figure~\ref{f:Q_mb19_5}.  We see configuration dependence at all scales, and the 
V-shape becomes more prevalent on larger scales; behavior that is reflected in both color 
sub-samples.  The red and blue galaxies show a larger disparity at smaller scales, 
with a significant increase in the clustering difference at $r_1 = 3 \hmpc$ for both 
the redshift space and projected $Q(\theta)$.  
Unlike the luminosity dependence, we notice a significant change in the configuration 
dependence between the color samples.  This effect is most noticeable in the $r_1 = 3 
\hmpc$ configuration for the projected $Q(\theta)$, which shows the red galaxies with a 
stronger configuration dependence than either the blue or full samples.  At configurations 
where $r_1 \ge 6 \hmpc$, the blue sub-sample shows greater configuration dependence than 
the red, which can be best seen in redshift space.  Overall, red galaxies typically show 
larger values of $Q(\theta)$ than the blue and full samples. 

\begin{figure*}
  \includegraphics[angle=270,width=\textwidth]{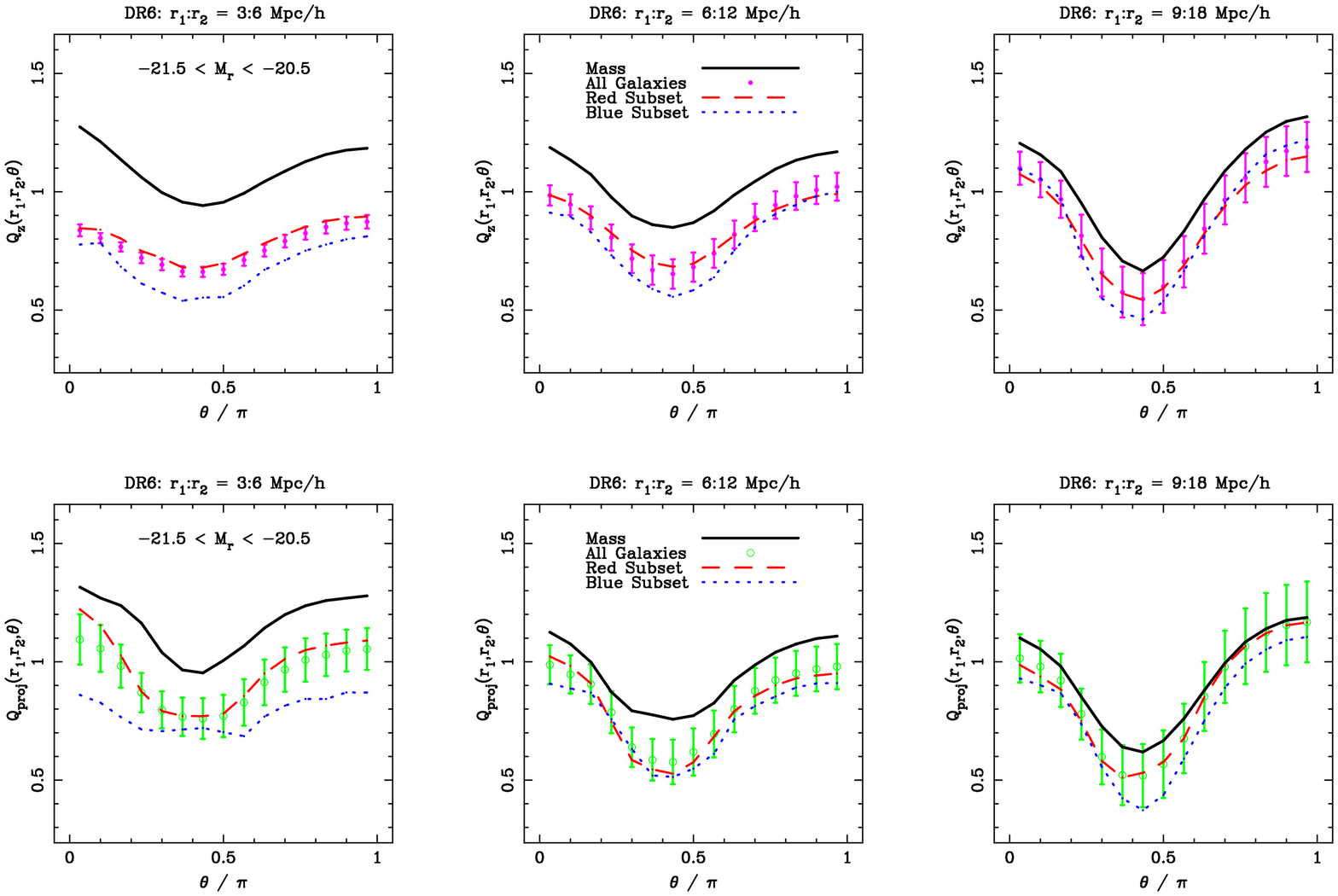}
  \caption[Reduced 3PCF of the $-21.5 < M_r < -20.5$ DR6 galaxy sample with color separation]{
    We show the configuration dependence of the reduced 3PCF, as in Figure~\ref{f:Q_mb19_5}, 
    but for the brighter LSTAR galaxy sample defined by $ -21.5 < M_r < -20.5 $.  
    The sample is divided into ``red'' (red, dashed line) and ``blue'' (blue, dotted line) 
    sub-samples based on SDSS $g-r$ color (see sample description for details).  
    The top row consists of redshift space measurements, and the bottom row depicts projected measurements.  
    The three columns correspond to different scales specified by the first side of the triangle ($r_1$) 
    representing the smallest scale measured.
    The black line denotes measurements on dark matter particles from the Hubble Volume simulation, 
    but matching the selection of the galaxies and includes redshift distortions.
    Error bars denote $1$$\sigma$ uncertainties calculated from $30$ jackknife samples.
  } \label{f:Q_mb20_5}
\end{figure*}

We show the reduced 3PCF for our LSTAR sample ($-21.5 < M_r < -20.5$) in 
Figure~\ref{f:Q_mb20_5}.  We still see configuration dependence at all scales, including  
color sub-samples --- but the color dependence appears diminished with respect to 
the FAINT sample in Figure~\ref{f:Q_mb19_5}. Again, the red galaxies at $r_1 = 3 \hmpc$ show a 
stronger configuration dependence than the blue according to $Q_{proj}$.  This sample 
has a volume almost four times larger than the FAINT sample, and the large scale 
$r_1 = 9 \hmpc$ measurements show significantly smaller uncertainties.  As we speculated 
before, the blue galaxies appear to have a slightly stronger configuration dependence.

Although the difference between red and blue sub-samples is both small and within the 
reported uncertainties, it might be significant as the errors are dominated by cosmic 
variance.  If the blue sub-sample does have a stronger configuration dependence than the 
red sub-sample, it would support the typically understood notion that blue galaxies more 
commonly populate the filamentary structures at these scales.  

On smaller scales ($r_1 = 3 \hmpc$), the blue sub-sample in projected space 
shows less configuration dependence than the red in both samples. We attribute this to a 
real clustering difference that is obscured by redshift distortions in the redshift space 
measurements. However, both $Q_z$ and $Q_{proj}$ show differences in the amplitude between 
``red'' and ``blue'' populations at the smaller triangle scales, but this difference 
declines at the larger scales.

\subsubsection{Comparing Galaxy and Mass Clustering}
\label{ss:Q_mass}

In Figures~\ref{f:Q_mb19_5} and \ref{f:Q_mb20_5}, we also plot measurements of the 
reduced 3PCF based on the dark matter particles in the Hubble Volume (HV) simulation.
Using \nbody\ simulations enables reliable predictions well into the non-linear regime 
where accurate analytic models do not exist \citep[although some recent work has been 
introduced, see][]{smith:08}.  
In addition, we can include observational systematics by trimming HV particles to match 
the exact selection and volume of the observed galaxy sample, and include the effects of 
redshift distortions.  The HV measurement serves as a comparison between clustering of 
the observed galaxies and that expected from gravitational evolution of a \lcdm\ mass field. 
Examining the figures, we make two observations: (1) the general shape is similar 
to that of the galaxy reduced 3PCF, and (2) there is a significant offset in amplitude 
between galaxies and mass.

\begin{figure*}
  \includegraphics[angle=270,width=\textwidth]{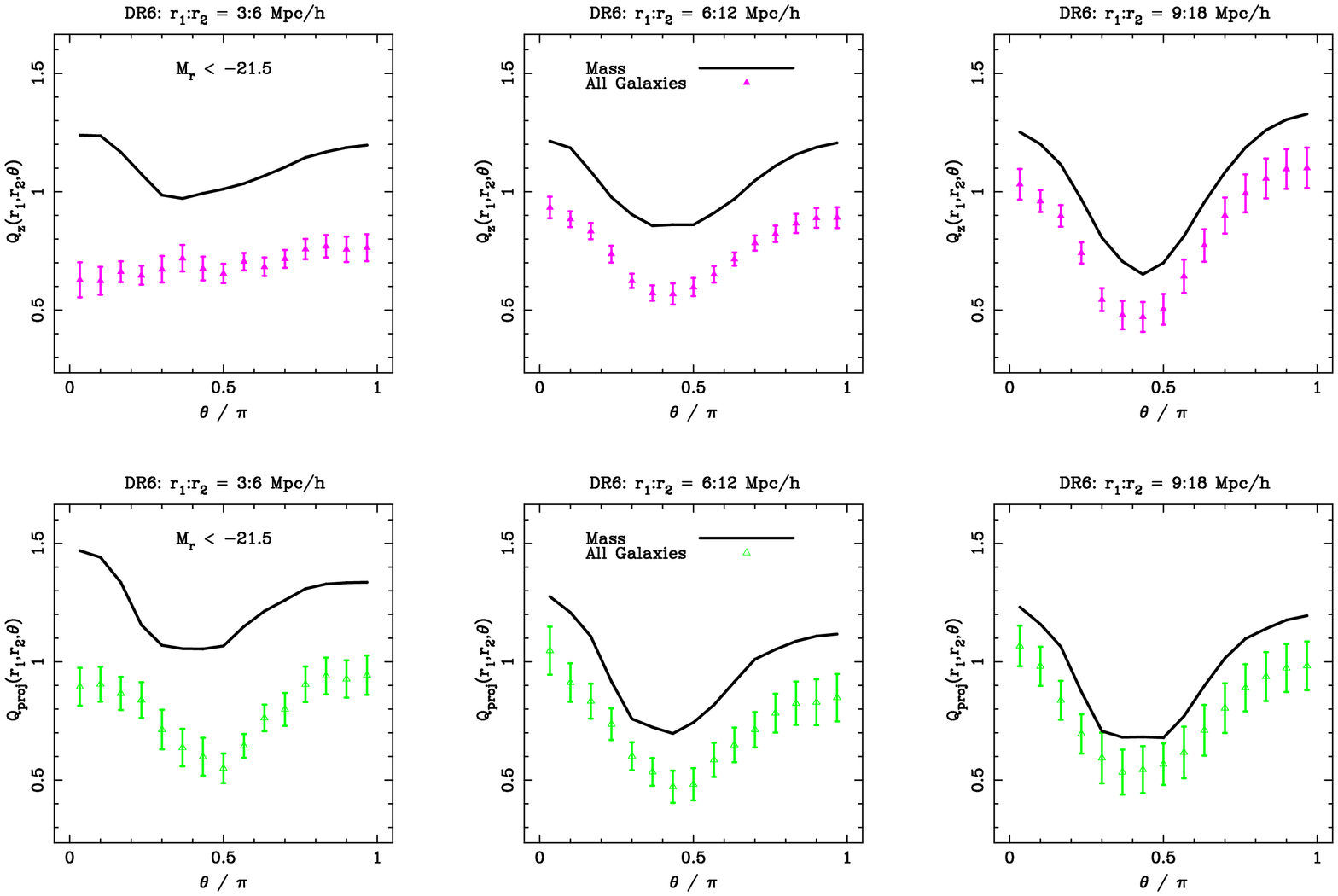}
  \caption[Reduced 3PCF of $M_r < -21.5$ DR6 galaxy sample]{
    We show the configuration dependence of the reduced 3PCF of SDSS DR6 galaxies with 
    $M_r < -21.5$.  The top row consists of redshift space measurements, and the bottom 
    row depicts projected measurements.  The three columns correspond to different scales 
    specified by the first side of the triangle ($r_1$) which represents the smallest 
    scale measured.  The black line denotes measurements on dark matter in the 
    Hubble Volume simulation, matching the selection of the galaxies and includes redshift 
    distortions.
    Error bars denote $1$$\sigma$ uncertainties calculated from $30$ jackknife samples.
  } \label{f:Q_mt21_5}
\end{figure*}

We show our measurement of the reduced 3PCF $Q$ on our BRIGHT galaxy sample 
($M_r < -21.5$) in Figure~\ref{f:Q_mt21_5}. As shown previously, the black line depicts 
measurements from the HV simulation in comparison to the symbols that denote the galaxy 
measurements.  The BRIGHT galaxy sample covers the largest volume, but is a factor of 
10 less dense than the LSTAR sample and almost $30$ times less dense than FAINT.  

It remains clear that galaxies do not cluster exactly the same as the simulated mass 
field.  However, there are significant commonalities.  The overall shape is the same, 
and the predominant effect is an offset.  
Remember, the nature of the reduced 3PCF makes it insensitive to cosmology 
(excepting the slope of the power spectrum), so the differences we observe in $Q$ are 
unlikely to be due to variation in assumed cosmology by the HV simulation.
There is a large volume of work showing that 
galaxies are known to be biased tracers of the mass field, which offers a natural 
explanation of the offset \citep{cooray:02}. The luminosity dependence of the reduced 
3PCF discussed in \S\ref{ss:Q_lum} is further evidence of the same.  The difference 
between clustering of galaxies and the underlying mass distribution is commonly referred 
to as galaxy-mass bias and depends on galaxy properties such as luminosity.  We constrain 
the non-linear galaxy-mass bias parameters relating to these measurements in a companion 
paper (McBride et al. 2010). 

\subsection{Covariance of Galaxy Samples}
\label{ss:covar}

Correctly accounting for the covariance in data measurements is essential to accurately 
constrain theoretical models. If significant covariance exists, neglecting it can result 
in a statistical bias (inaccurate constraints) or an overestimate of significance. 
However, proper estimation and use of the covariance matrix remains tricky.  
For statistical measures of clustering in LSS, the covariance matrix encodes higher-order 
information.  The covariance matrix of the 2PCF includes significant third and 
fourth order terms, whereas the covariance of the 3PCF has leading order contributions 
from up to sixth order.  The covariance matrix itself is a complementary measure of 
clustering.  

We typically use the inverse of the covariance matrix to constrain models, such as 
a common $\chi^2$ determination.  Doing this makes an analysis extremely sensitive to 
the noise properties of the covariance matrix (the poorest determined eigenmodes can have a 
dramatic impact on the $\chi^2$ value).  While this fact is often overlooked in LSS 
analyses, ways of accounting for poorly resolved modes are well established for the 
2PCF \citep[see e.g.][]{norberg:09} and 3PCF \citep[see e.g.][]{GS05}.  If a measured 
statistic is poorly resolved, i.e. the dominant contribution to the error is shot-noise, 
the off-diagonal elements of the covariance matrix can become excessively noisy. 
In such a case, using the full covariance matrix might be a poorer choice than 
a diagonal approximation \citep[demonstrated in][]{chen:06}.

For our galaxy samples, we resolve the covariance matrix using jackknife re-sampling 
methods defined by $30$ equal-area regions on the sky (a factor of two over the number 
of measured bins).

\begin{figure*}
  \includegraphics[angle=270,width=\textwidth]{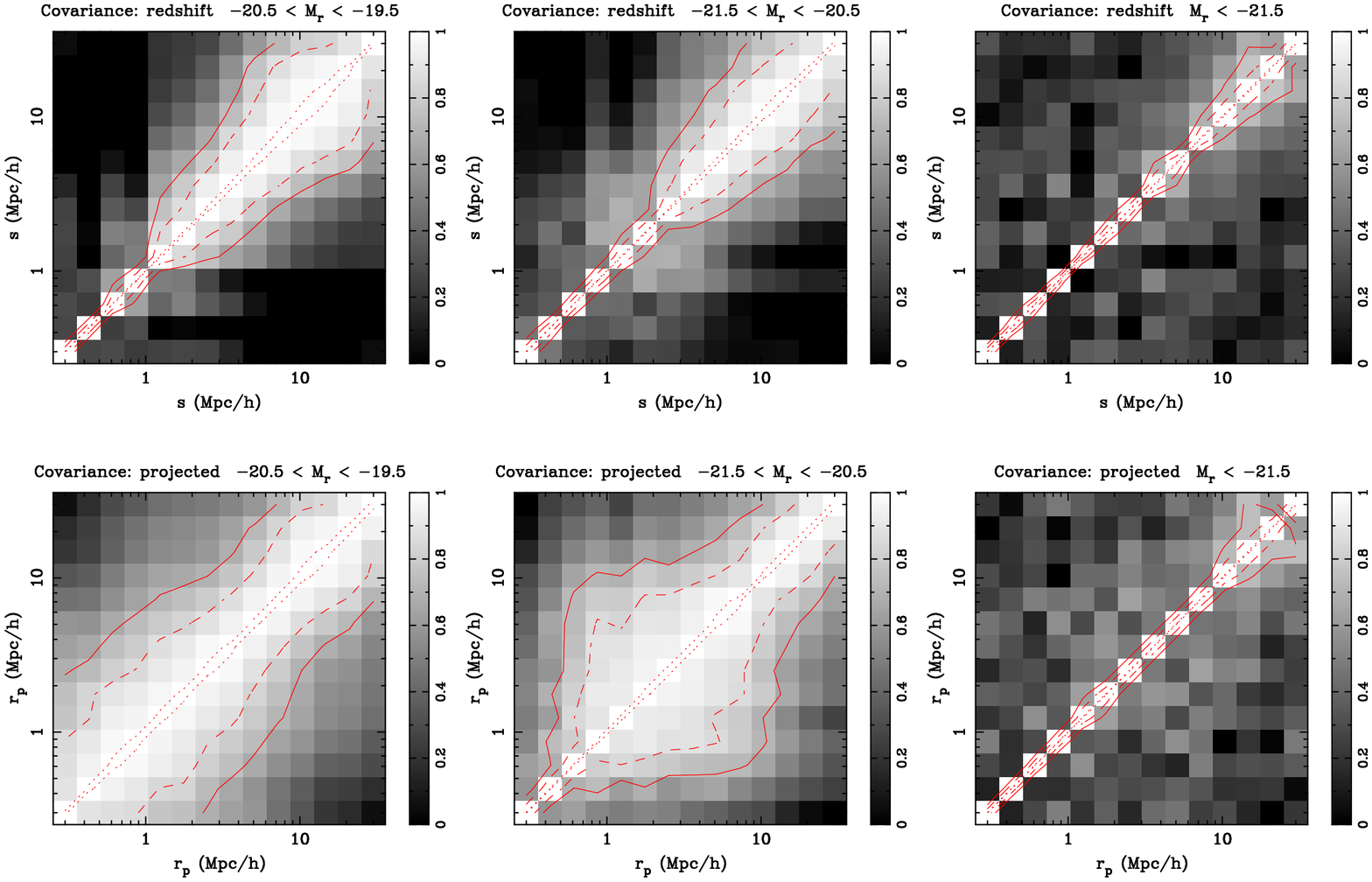}
  \caption[2PCF covariance matrices for SDSS galaxies]{ The normalized covariance matrix for the 2PCF,  
  both in redshift (top row) and projected space (bottom row) for our three DR6 galaxy samples of 
  different magnitudes.  From left to right, the panels indicate FAINT to BRIGHT galaxy samples.
  The matrix is normalized such that diagonal elements are set to unity, rather than the $1$$\sigma$ error.  
  The solid, dashed and dotted contours correspond respectively to values of 0.70, 0.85 and 0.99.}
  \label{f:covar_2pcf}
\end{figure*}

We show the normalized covariance matrix of the 2PCF for our three galaxy samples in 
Figure~\ref{f:covar_2pcf}.  At these scales we expect significant correlation in the covariance
matrix for all the 2PCF measurements.  In the power spectrum we expect the Fourier modes 
to be statistically independent \citep{bardeen:86} and the off-diagonal elements of 
the covariance matrix in the 2PCF preserve this Fourier space property.  We note the 
BRIGHT sample (right column) appears almost diagonal.  Since we still expect correlated bins for 
the sample, we interpret this result as an undersampled covariance matrix. This can occur when 
the most significant contribution is shot-noise (since Poisson errors are not correlated).  While 
this is still a ``true'' measure of the errors, care must be taken when using the full covariance 
matrix as the off-diagonal elements can be noisy.
For the two fainter samples, we see the projected $w_p(r_p)$ measurements appear much more correlated 
than the redshift space $\xi(s)$.  Since $w_p$ projects along the line-of-sight, the projection 
mixes scales between ($s = r_p$) and ($ s = (\pi^2 + r_p^2)^{1/2}$), inducing a correlation in the 
measurements.  Seeing this effect supports the validity our of covariance estimation.  Finally, 
we note that the correlation appears to increase for the fainter galaxy sample.  In the 2PCF, 
this increase of correlation with fainter galaxy samples has been previously seen
\citep{zehavi:05}.

\begin{figure*}
  \includegraphics[angle=270,width=\textwidth]{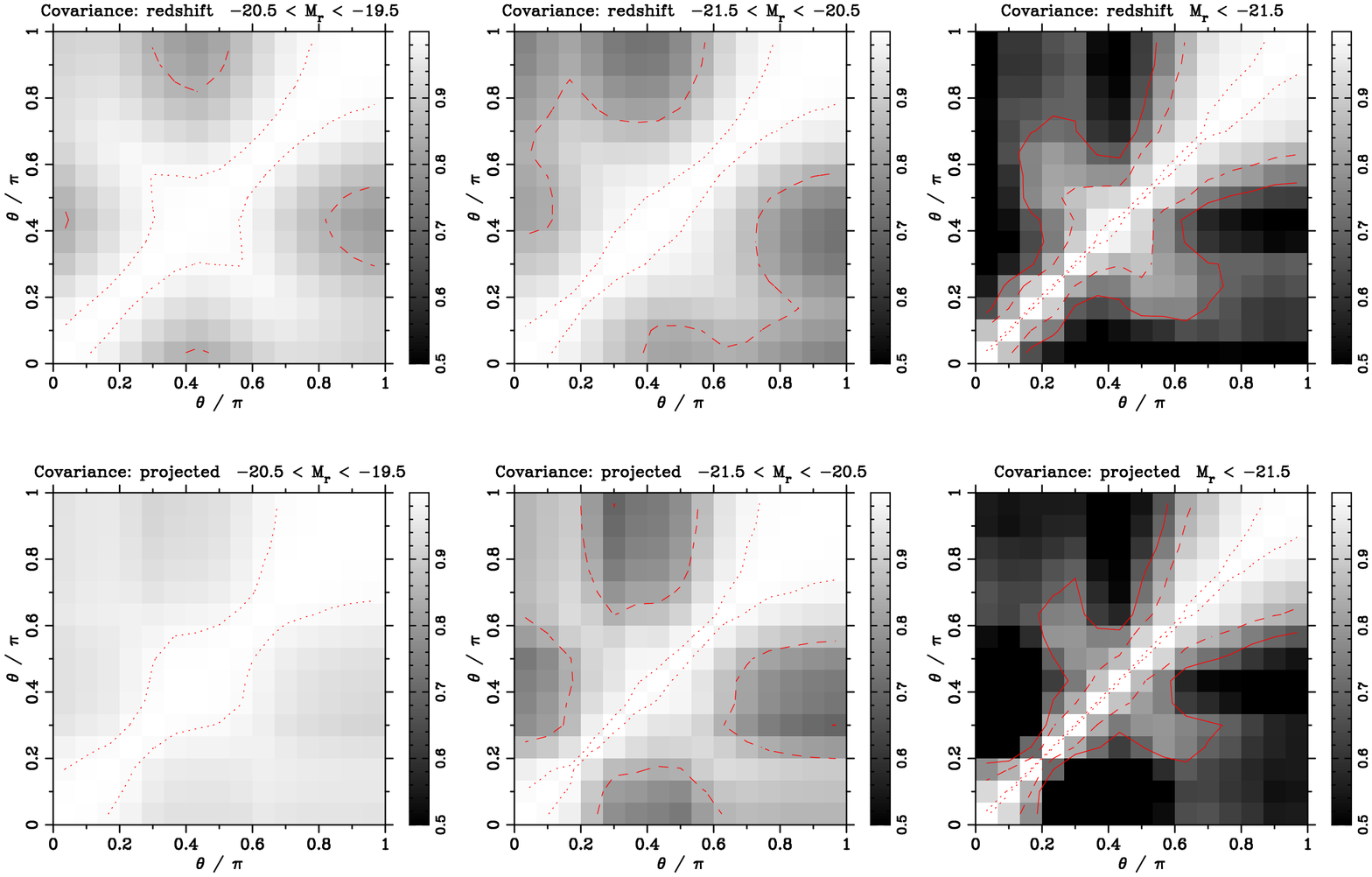}
  \caption[Reduced 3PCF covariance matrices for SDSS galaxies]{ 
  The normalized covariance matrix for the reduced 3PCF, both in redshift (top row) and 
  projected space (bottom row) for our three DR6 galaxy samples of different magnitudes.
  From left to right, the panels indicate FAINT to BRIGHT galaxy samples.
  The matrix is normalized such that diagonal elements are set to unity, rather than the $1$$sigma$ error.  
  This shows only the covariance for the largest triangles where $r_1 = 9 \hmpc$.
  The solid, dashed and dotted contours correspond respectively to values of 0.70, 0.85 and 0.99.}
  \label{f:covar_Qs9}
\end{figure*}

The covariance matrices for the reduced 3PCF on large scales ($r_1 = 9 \hmpc$) for the three 
galaxy samples are shown in Figure~\ref{f:covar_Qs9}. 
Note the significant off-diagonal structure for all samples, which is comparable to that seen 
in theoretical studies \citep[see figure 9 in][]{GS05}.  
The correlation between a few neighboring bins at $\theta \approx \pi$ is enhanced due to 
physical overlap of our binning scheme (the same triplet can be counted more than once).  
We again see that the correlation in the projected measurements indicate more correlation
in comparison to redshift space. Analogous to the 2PCF, we find that fainter galaxy samples 
appear to have increased correlation between bins of the reduced 3PCF. Although we show only the 
largest measurements, we find significant off-diagonal elements at all scales.  Clearly the use 
of a diagonal approximation to the covariance matrix would be a poor assumption. 




\subsection{Effects of Super Structures}
\label{ss:sstruct}

Large coherent structures or ``\emph{super} structures'', such as the Sloan Great Wall 
\citep[SGW;][]{gott:05}, can dramatically affect clustering measurements.  Detailed 
analyses on SDSS galaxy samples have documented this in both the 2PCF 
\citep{zehavi:02,zehavi:05, zehavi:10} and redshift space reduced 3PCF \citep{nichol:06}. 
One advantage of using jackknife re-sampling methods for error analysis is that we probe the 
variation between different spatial regions essentially ``for free''.
We investigate this variation using our $30$ jackknife samples.  There are several reasons 
this variation of the 3PCF is worth investigation beyond that in the literature.  First, our 
samples are based on a newer SDSS sample (DR6) which includes additional regions of the sky 
resulting in a larger volume than previous studies. Second, as we have repeatedly mentioned, 
measurements of the reduced 3PCF are affected by a chosen binning scheme making it important 
to document the effects of structures given our exact parameterization.  Finally, we want 
to understand the impact on detailed measurements of the projected reduced 3PCF for which 
the effect of super structures has not been previously investigated.

\begin{figure}
  \centerline{
    \includegraphics[angle=0,width=\linewidth]{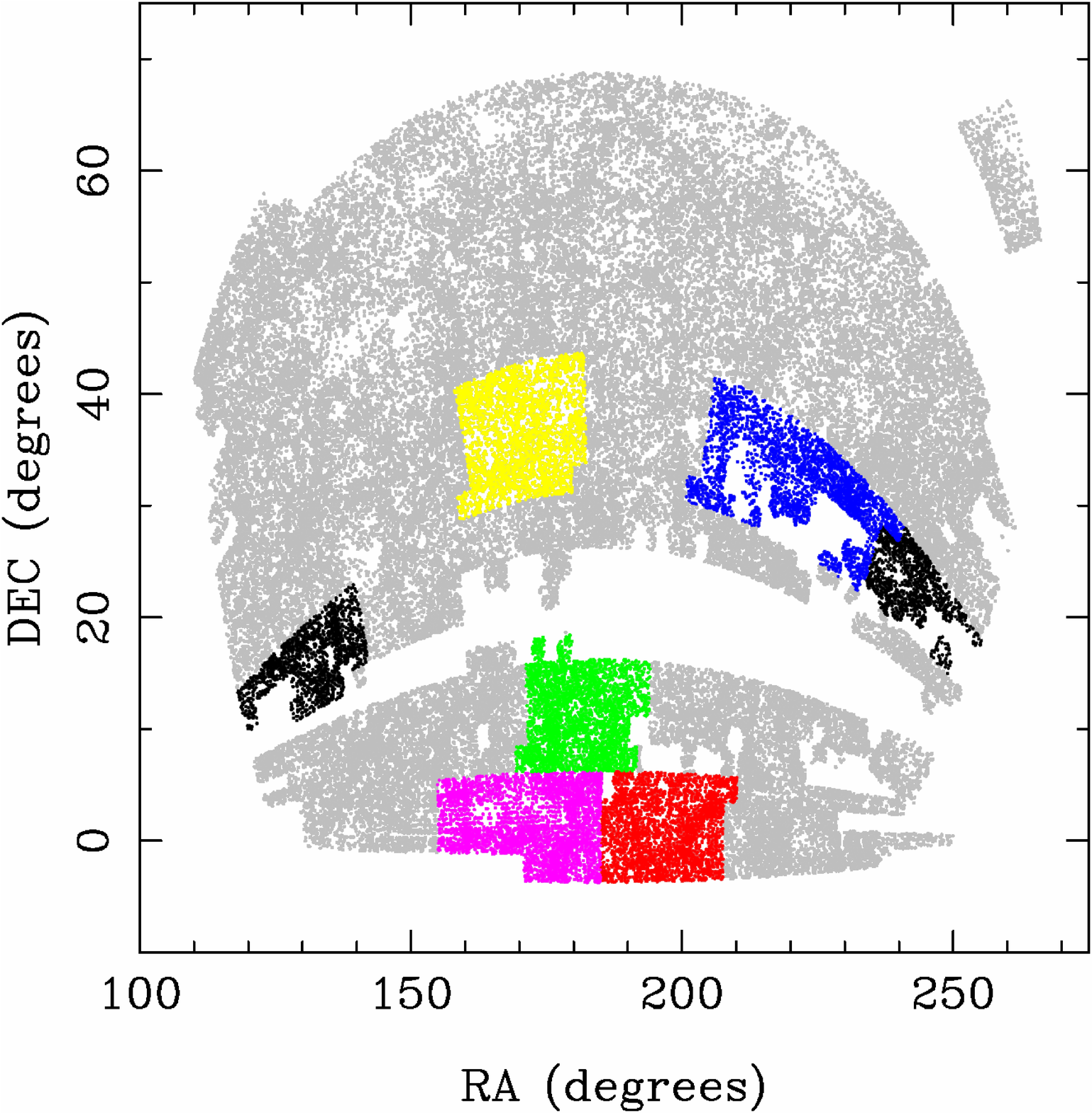}
  }
  \caption[SDSS North: select jackknife regions]{ 
    We show six selected jackknife regions on the sky (yellow, blue, black, green, 
    magenta, and red) in comparison with the full galaxy sample (grey points) for the SDSS 
    DR6 North footprint in J2000 equatorial coordinates.  The Sloan Great Wall \citep{gott:05} 
    is located at $\sim0$ in declination, and is included in two of the six selected 
    jackknife regions (red and magenta).
  }
  \label{f:jacksamples}
\end{figure}

We identify $6$ out of $30$ regions of the sky which show large deviations in the reduced 
3PCF.  These ``jackknife regions'' characterize a jackknife sample by being \emph{omitted} 
from a clustering measurement.  The northern SDSS DR6 footprint in shown in 
Figure~\ref{f:jacksamples}, with the entire sample displayed in gray and the six regions 
highlighted by color.  Two regions encapsulate the majority of the SGW, specifically the 
red and magenta regions at a J2000 declination of zero.  Overall, the jackknife regions 
appear contiguous and rectangular (excepting the apparent
geometry of the Aitoff sky projection). Please note the black region, however, which is 
split between two sides of the survey. The algorithm we use to define the jackknife 
regions must occasionally make such divisions.

Before examining the reduced 3PCF for these regions, let us briefly review the jackknife re-sampling 
method.  We excise a jackknife region from the full sample and measure the clustering.  This 
means that a measurement on a specific jackknife sample represents the clustering of the entire 
sample \emph{omitting} the jackknife region.  If the clustering on the jackknife sample deviates 
strongly from the average of all samples, it means that a specific jackknife region 
dominates the measurement for the entire sample.  Without that region, the overall clustering 
would be significantly different.  This is a profound concept, as we do not expect such 
clustering differences given the volume of these SDSS samples.

We investigate the effects on the reduced 3PCF in two of our three galaxy samples.  Although the jackknife 
regions based on sky location are consistent across both, the redshift limits vary.  This 
results in a volume that is not identical between samples, although it does overlap.  We use our 
LSTAR ($-21.5 < M_r < -20.5$) and FAINT ($-20.5 < M_r < -19.5$) samples which both include the 
SGW at a mean redshift at $z \approx 0.08$.  

To highlight the clustering deviations between samples, we plot the 
\emph{residual} of the reduced 3PCF, as defined by \eqref{eq:del}.  We subtract the mean reduced 3PCF 
of the $30$ samples from each individual measurement, and normalize this difference by the 
\emph{jackknife variance}.  As noted in \eqref{eq:cov_jack}, the prefactor for the jackknife 
variance is ($\frac{N - 1}{N}$), and not the familiar $\frac{1}{N}$, so error estimates 
do not decrease with $N$.  In terms of the residuals, this means no jackknife sample will 
deviate significantly beyond the error. This scaling of the residuals emphasizes how each jackknife 
region significantly contributes to the $1\sigma$ error estimate.

\begin{figure}
  \centerline{
    \includegraphics[angle=0,width=1.0\linewidth]{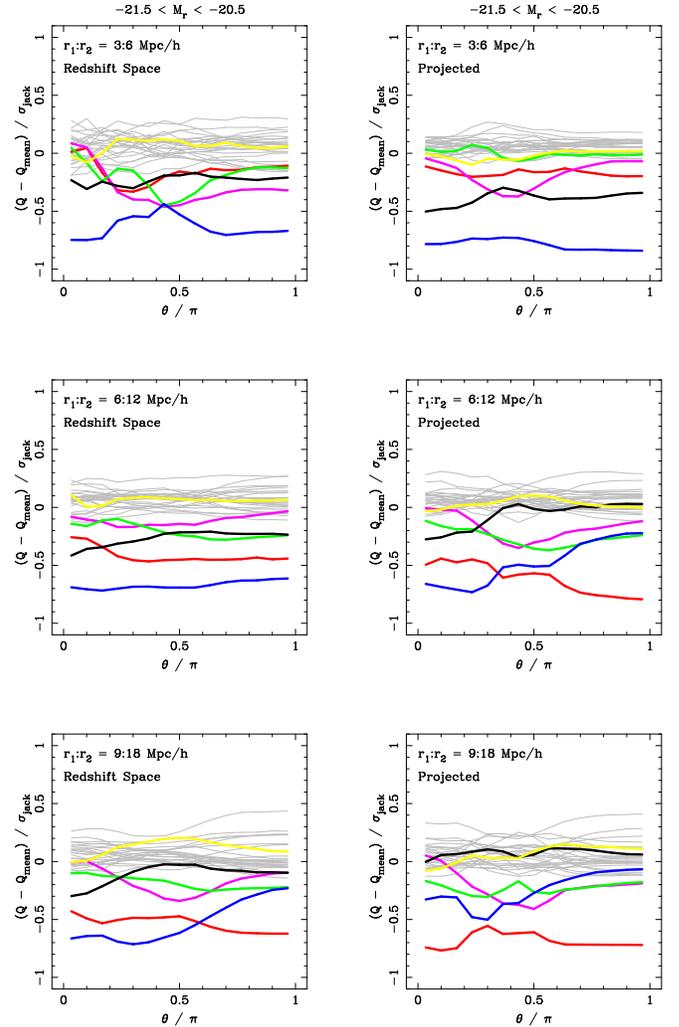}
  }
  \caption[Reduced 3PCF residuals for $-21.5 < M_r < -20.5$]{ 
    The residuals, as defined in \eqref{eq:del}, for 30 jackknife samples of the reduced 
    3PCF in redshift and projected space for the DR6 galaxies with $-21.5 < M_r < -20.5$.  
    A jackknife sample is defined by taking the full sample and excluding a specific 
    jackknife region.  Colors of the six selected jackknife samples correspond to 
    excluding galaxies in the region of matching color in Figure~\ref{f:jacksamples}. 
  } \label{f:jacktest_vl_mb20_5}
\end{figure}

Figure~\ref{f:jacktest_vl_mb20_5} shows the residual of the reduced 3PCF for the LSTAR 
galaxy sample ($-21.5 < M_r < -20.5$).  The left three panels are results for redshift space, 
$Q_z$, and the right correspond to projected space, $Q_{proj}$.  The scale increases downward 
in order with $r_1 = 3,6$ and $9 \hmpc$.  The gray lines represent results from the $30 - 6 = 24$ 
``ordinary'' jackknife samples.  The colored lines correspond to jackknife samples that omit the 
jackknife regions highlighted in Figure~\ref{f:jacksamples}.  Most of the regions are close to the 
mean value (i.e. around zero), but several samples deviate, which 
always happens in the negative direction.  If a structure in the jackknife region, which is 
included in \emph{all} other samples, boosts $Q(\theta)$, then a negative residual results.  In the 
LSTAR sample, $Q_z$ appears to be boosted at all scales by galaxies in the blue jackknife region.  
As scale increases, structure in the red region also appears to increase $Q_z$.  Both regions also 
have an effect on $Q_{proj}$, with red clearly dominating at $r_1 = 9 \hmpc$ and blue at the 
small scale $r_1 = 3 \hmpc$ measurement.  In $Q_{proj}$, we note the black region, which is a 
physical neighbor to the blue region, strongly affects the small scale reduced 3PCF.

It appears the red region, which encloses a significant portion of the SGW, predominantly affects 
the reduced 3PCF on large scales; this has been seen before in the reduced 3PCF by \citet{nichol:06}.  
The blue and neighboring black regions were first included in the DR5 release.  At smaller scales, these 
regions clearly dominate the reduced 3PCF measurement.  
We visually inspect the galaxy distribution within the several of the anomalous regions, 
and find that these jackknife regions (consisting of $1/30$ of the entire SDSS area) 
enclose a single dominant super structure, or sometimes just part of one structure 
straddled across several regions.
For example, there is an extremely clustered region at the boundary of the blue and black regions 
with a median redshift around $z \approx 0.11$.  This is clearly distinct from the SGW.  We refer 
to such regions as ``super'' structures since, like the SGW, they are coherent 
overdensities that are not gravitationally self-bound in contrast to a galaxy cluster.  Individually, 
these regions shift the reduced 3PCF by less than 10\% -- but this analysis neglects cumulative 
effects (remember, the same super structure is split between the blue and black regions).  

\begin{figure}
  \centerline{
    \includegraphics[angle=0,width=1.0\linewidth]{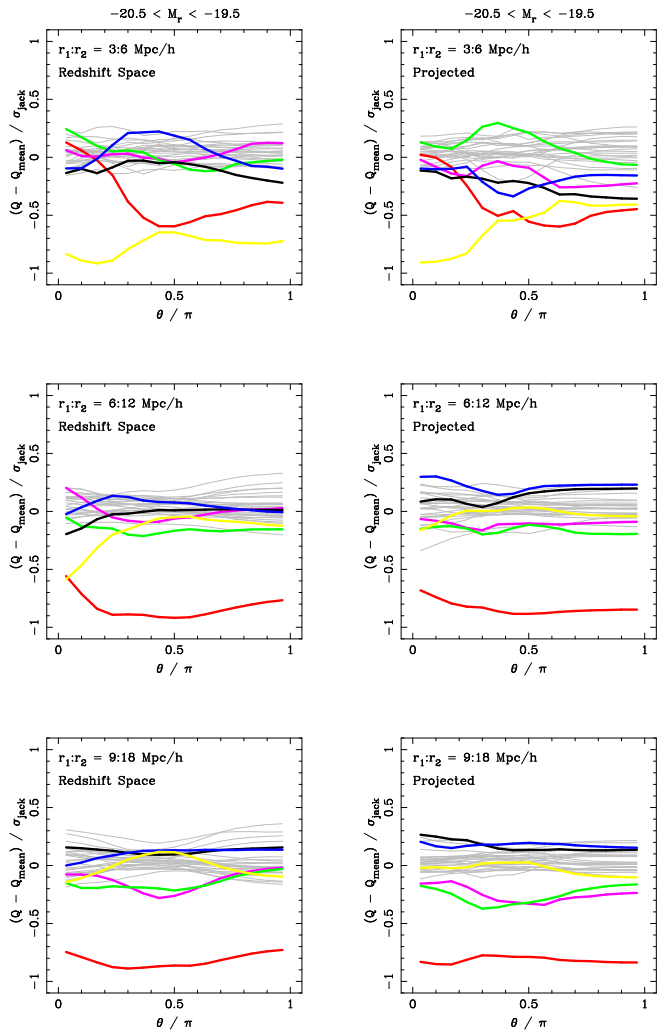}
  }
  \caption[Reduced 3PCF residuals for $-20.5 < M_r < -19.5$]{ 
    The analogous figure to \ref{f:jacktest_vl_mb20_5}: the residuals as per 
    \eqref{eq:del}, for 30 jackknife samples of the reduced 3PCF in redshift and projected 
    space for the DR6 galaxies with $-20.5 < M_r < -19.5$. A jackknife sample is defined 
    by taking the full sample and excluding a specific jackknife region.  Colors of the 
    six selected jackknife samples correspond to excluding galaxies in the region of 
    matching color in Figure~\ref{f:jacksamples}. 
  } \label{f:jacktest_vl_mb19_5}
\end{figure}

We examine the FAINT galaxy sample ($-20.5 < M_r < -19.5$) in Figure~\ref{f:jacktest_vl_mb19_5}.  
Recall from the reduced 3PCF measurement in Figure~\ref{f:Q_mb19_5}, that the uncertainties were 
quite large, much more than the LSTAR sample.  The residuals at the large scales 
are completely swamped from the clustering in the red region.  Excluding this region 
changes the reduced 3PCF by approximately $20\%$, which we attribute to galaxies clustered in the SGW.  
The black and blue regions seem inconsequential for this sample; this makes sense as this 
fainter sample has a maximum redshift of $z = 0.086$, thereby trimming the structure at 
$z \approx 0.11$ which accounted for the significant deviation in the LSTAR sample.  
We also see that a new region, denoted in yellow, dominates at small scales.

\section{Discussion}
\label{s:discussion}

Our clustering measurements on our galaxy samples produced the following main results:
\begin{itemize}
  \item $Q(\theta)$ exhibits configuration dependence at all scales ($3-27 \hmpc$), which is 
    not consistent with $Q$ being simply a constant.
  \item Larger scales ($9-27\;\hmpc$) exhibit a stronger configuration dependence and display 
    the V-shape expected from gravitational evolution.
  \item All galaxy samples show significantly different clustering strength than the mass estimates.
  \item Uncertainties for the reduced 3PCF increase for larger scales and with smaller sample volume.
\end{itemize}
Generally, the shape of our SDSS galaxy measurements appear similar to predictions of gravitational 
collapse in the \lcdm\ model realized by the Hubble Volume simulation. The difference in clustering strength 
between galaxies and mass predictions can be accounted for as galaxies of these luminosities are
expected to be biased tracers of the mass field.  We constrain the non-linear galaxy-mass bias 
using a detailed error analysis in a companion paper (McBride et al. 2010).

We studied measurements of three galaxy samples characterized by different luminosities, 
attributing observed variations as a luminosity dependence in the reduced 3PCF.  In 
\S\ref{ss:Q_lum} we found: 
\begin{itemize}
  \item Luminosity dependence affects all scales at about the same level, which primarily 
    changes the amplitude of $Q$. 
  \item The discrepancy between galaxies and mass predictions appears larger for brighter galaxies.
  \item The covariance matrix appears more correlated for fainter galaxy samples.
\end{itemize}
The division of two volume-limited samples into sub-populations based on color led to 
these observations:
\begin{itemize}
  \item The $g-r$ color split isolates populations with different clustering properties, 
  \item Red sub-samples have a higher average $Q$ value than blue.
  \item The difference between red and blue sub-samples is greatest at small scales and 
    includes changes to the configuration dependence. 
  \item The effects of galaxy color on $Q$ are more pronounced than luminosity at small scales, 
    where galaxy color more dramatically affects the configuration dependence.
\end{itemize}

A concern might arise that the defined samples are not complete for both color 
definitions. This could result from possible evolutionary effects \citep[discussed in the 
context of DEEP2 in][]{gerke:07} or inaccuracies in the $K$-correction due to differences in galaxy
morphology.  We confirmed that there were minimal to no differences in the measured reduced 3PCF 
by creating samples defined to be more conservative in their sample description, where we found 
the largest variation to be less than 10\% of the reported errors. 

An analysis of the 2PCF for SDSS galaxies \citep{zehavi:05, zehavi:10} conceptually explains the 
clustering difference of the brighter galaxy samples. Brighter galaxies remain ``more biased'' than 
fainter samples resulting in a larger discrepancy between galaxies and mass for brighter 
galaxy samples.  We see a similar result in our clustering measurements and find that $Q$ shows less 
discrepancy between samples since $Q \propto 1/b$ whereas $\xi \propto b^2$ (``more biased'' refers 
to a higher linear bias value of $b$). The stronger effect of color separation on smaller scales 
($3-9\hmpc$) agrees with the conclusions of \citet{blanton:05} that galaxy color is strongly tied to 
local environment.  Recent analysis of the 2PCF by \citet{zehavi:10} showed striking 
differences in clustering between luminosity and color, where luminosity predominately 
shifts the amplitude, whereas different color samples show a different slope of the 2PCF.  We find
that the configuration dependence of $Q_{proj}$ is stronger for red sub-samples at smaller scales 
($r_1 = 3 \hmpc$).  This scale is still beyond a reasonable size for individual dark matter halos.  
However, it might suggest that regions outside large halos have a surrounding anisotropic distribution 
of red galaxies, perhaps tracing the infall regions.  This interpretation supports 
observations of the angular distribution of satellite galaxies by \citet{azzaro:07}, where they 
find an alignment of satellites along the major axis of their host. They find the strongest evidence 
of alignment when considering red satellites of red hosts, which would be best represented by our red 
sub-sample.
On measurements of large scale triangles ($r_1 = 9 \hmpc$), we conclude that both red and blue populations 
exhibit similar configuration dependence given the large errors. However, we speculate that the blue 
galaxies occupy regions that show stronger shape dependence.

A detailed study of the reduced 3PCF in redshift space \citep{gaztanaga:05} was conducted on a different 
redshift survey: the two-degree field galaxy redshift survey \citep[2dFGRS;][]{2dFGRS}.  Their 
parameterization of triplets is very close to ours, and the analysis is in qualitative agreement 
with our measurements of $Q_z$.  They also divided a volume-limited sample into ``red'' and ``blue'' 
sub-samples, noting that red galaxies typically have a larger value of $Q_z$ below $12 \hmpc$.  

The reduced 3PCF of SDSS galaxies was also studied in redshift space by \citet{kayo:04}.  Their
measurements that are sensitive to shape (see their Figure~13) show a configuration dependence that
grows with scale.  Their $Q_z$ for $s=2.5 \hmpc$ shows very little configuration dependence with
$Q_z \leq 1$ for all $\theta$, which is comparable to our measurements at $s=3.0 \hmpc$, and confirm
the effects of redshift distortions on these scales.  In contrast to their findings, our measurements
suggest a luminosity dependence of the reduced 3PCF which is apparent in the equilateral $Q_{eq}$
(Figure~\ref{f:Qeq}) but more so in the configuration dependent $Q(\theta)$
(Figure~\ref{f:Q_lumcmp}).  However, even our study suggest a weak statistical significance,
especially on the larger scales.  The difference between our results and \citet{kayo:04} is likely
due to statistical strength of the galaxy samples -- since DR6 contains many more galaxies in a much
larger volume than their earlier data, it makes the weak luminosity dependence noticeable.  We note 
a discrepancy between the results with respect to color dependence. \citet{kayo:04} see very little 
clustering difference between populations on our small scales (middle panels of their Figure~13).  In addition, their
measurements show the blue population has larger value of $Q$ than the red on larger scales (bottom
panels of their same figure), whereas our results show little difference (our right hand panels of
figures~\ref{f:Q_mb20_5} and \ref{f:Q_mb19_5}).  However, the latter effect is within $1\sigma$
given the uncertainties of both studies, making any conclusions suggestive at best.  The differences
might be accounted for by one of the many subtle differences of the analyses, which include 
(1) different SDSS dataset, (2) different galaxy sample and color definitions, and (3) different
choices of triangle parameterization and binning.  Sorting out these systematic differences would 
require a detailed joint comparison which is not warranted by the low statistical significance of 
the discrepancy.

We compare clustering of the reduced 3PCF in redshift and projected space to find:
\begin{itemize}
  \item $Q_{proj}$ successfully recovers configuration dependence at small scales ($3-9 \hmpc$), 
    which is lost in $Q_z$.
  \item $Q_{proj}$ has a higher amplitude at small scales than $Q_z$.
  \item $Q_z$ and $Q_{proj}$ both converge at large scales ($9-27 \hmpc$).
\end{itemize}
Both $Q_z$ and $Q_{proj}$ were measured by \cite{jing:04}.  Although they analyze entirely 
different galaxy data (2dFGRS), we expect similar behavior between redshift and projected space 
measurements. Their results do not agree completely with our findings on comparable scales and 
triplet configuration. While they notice an increase in the amplitude of $Q$ between redshift and 
projected measurements in their ``full'' sample (compare their figures~7~and~12 for $r=3.25 \hmpc$ 
with $u = 2$), it is not as much of a difference as our results suggest.  They also see no configuration 
dependence in $Q_{proj}$.  We suspect their measurements obscure these features due to their choice 
of binning and triangle parameterization, as described in \cite{GS05}.  A separate analysis of 2dFGRS 
data by \cite{gaztanaga:05}, which uses a parameterization similar to what we employ, presents $Q_z$ 
measurements that show more configuration dependence than \cite{jing:04}. 

We have seen that the covariance of the reduced 3PCF yields significant structure at all resolved scales and 
in all of our galaxy samples.  While our choice of wide bins contribute to the correlation, 
especially when $\theta$ approaches zero or $\pi$, we conclude most of the correlation is physical.  
Measuring the configuration dependence of the reduced 3PCF requires closely packed measurements.  
The range of scales probed by a specific $Q(r_1, r_2, \theta)$ measurement is the change in scale of 
the third side, equal to $2 r_1$ for our measurements.  Empirically estimating the covariance 
enables both systematic correlations (i.e. overlapping bins) and physical covariance between 
reduced 3PCF values to be taken into account for quantitative constraints.  There is a limit to 
the effectiveness of empirical determinations: if we bin the correlation function too finely, such 
that pair and triplet counts become poorly sampled, the uncertainties will be dominated by Poisson 
noise and the covariance matrix will look diagonal. This would be a false representation of the 
correlation, and care must be taken to properly account for noise in the covariance matrix if used 
in an analysis.  We note that we approached this undersampled limit for the covariance of our 
brightest sample ($M_r < -21.5$) even with our wide binning scheme (bin-width was $0.25 r$, 
where $r$ is the scale probed).

If we restrict ourselves to the quasi-linear regime where $\xi < 1$, represented in the 
reduced 3PCF by the largest scale measured ($r_1 = 9 \hmpc$), we see significant structure to the 
covariance matrix. This type of structure, almost an ``X'' pattern, has been seen in 
simulations \citep[e.g.][]{GS05}.  We notice that the overall correlation increases as the 
galaxy sample becomes fainter for this scale.  This luminosity dependence of covariance 
has been seen in the projected 2PCF \citep{zehavi:05}. This work is the first time it has 
been resolved in observational measurements of the reduced 3PCF, which we note in both 
redshift and projected space.  An important implication of this result is that care must be 
taken to properly estimate the covariance matrix specific to the galaxy sample being studied.
The correlation matrix of one galaxy sample is not necessarily representative of another. 

We investigate the variation of clustering in jackknife samples to enable a view of clustering 
in SDSS galaxy samples with an alternative perspective.  A few regions, which each likely contain 
one or perhaps two ``super structures'', dictate the clustering of the entire sample -- even with 
the sizable volume represented by the SDSS samples. Different structures affect different scales. 
We clearly showed anomalous structures in a few rare regions dictate the accuracy of clustering 
measurements. The standard proposed solution is to continue collecting bigger samples in the hope 
of a large enough volume to average over multiple rare ``super'' structures.  Clearly, these super 
structures can affect quantitative descriptions of bias.  How best to handle this in detailed 
analyses remains an open question, but is most often addressed by comparing sub-samples with 
matching volumes (so a specific structure affects them equally).

We agree with \citet{nichol:06} that the reduced 3PCF shows more sensitivity to the presence 
of super structures than lower order statistics such as the 2PCF \citep{zehavi:05,zehavi:10}.  
Although we do not include details in this paper, we also noticed marked differences in the 2PCF due 
to our jackknife regions. With the reduced 3PCF, deviations caused by the super structures appear largely 
as an amplitude offset affecting triplets of all configurations equally.  It would be interesting to 
see if the lack of configuration dependence in the deviations between jackknife regions continues 
for smaller scales that approach the size of large halos.  The configuration dependence at the scales 
we measure are predominately due to filamentary structures ($3-27 \hmpc$ correlates structure 
between 3 different halos), making it unlikely that one specific region would dramatically enhance 
or erase this signature. 

\section{Summary} 
\label{s:summary}

In this paper we present clustering measurements of three volume-limited samples of SDSS galaxies.  
We investigate the 2PCF in \S\ref{ss:2pcf} and find clustering measurements consistent with other 
analyses of SDSS data \citep{zehavi:05,zehavi:10}. Specifically, we note the 2PCF is reasonably 
well approximated by a power-law model and brighter galaxies result in stronger clustering at 
all measured scales ($0.3-30 \hmpc$).

We consider the reduced 3PCF in \S\ref{ss:Qeq} and \S\ref{ss:Q_conf} and find significant 
configuration dependence on intermediate to large scales ($3-27 \hmpc$), in general agreement 
with predictions from \lcdm. These results are in contrast to the hierarchical ansatz where the 
reduced 3PCF shows no dependence on triplet shape.  Below $6 \hmpc$, the redshift space reduced 
3PCF shows reduced power and weak configuration dependence in comparison with projected measurements. 
These results indicate that redshift distortions, and not galaxy bias, can make the reduced 
3PCF appear consistent with the hierarchical ansatz. We address the luminosity dependence of our 
samples in \S\ref{ss:Q_lum} and color dependence in \S\ref{ss:Q_color}.  Compared to the lower 
order 2PCF, the reduced 3PCF exhibits a weaker dependence on luminosity with no significant 
dependence on scales above $9 \hmpc$.  On scales less than $9 \hmpc$, the reduced 3PCF 
shows a more dramatic dependence on galaxy color than on luminosity, which includes 
significant changes to the configuration dependence.

We resolve the covariance matrices of our clustering measurements in \S\ref{ss:covar}, 
calculated by jackknife re-sampling using $30$ samples.  We find significant structure in 
the covariance with large off-diagonal elements depicting strong correlations.  These
results demonstrate that an assumption of a diagonal covariance matrix is a poor choice,
and the correlations must be taken into account for any quantitative analysis. 
The covariance matrix can be improperly resolved, such as when measurement bins are 
too small, complicating the use of the full covariance matrix due to noisy modes.
We show that the overall correlation generally increased with fainter galaxy samples, 
suggesting a luminosity dependence to the structure of the covariance.
Clearly care must be taken to properly estimate the covariance matrix specific to 
the galaxy sample being used for quantitative constraints --- a fact which is often 
overlooked in recent work on the reduced 3PCF.

In \S\ref{ss:sstruct}, we demonstrate how large coherent structures, referred to as ``\emph{super}
structures'', affect these clustering measurements. We use $30$ independent regions on the 
sky, and show that $6$ of the $30$ produce anomalous deviations in clustering, with different 
structures dominant at different scales.  These regions, each of which contain one or perhaps 
two super structures, dictate the clustering of the entire sample -- even with the sizable volume 
of the SDSS galaxy samples.  Two of these regions are coincident with the huge structure known 
as the Sloan Great Wall \citep[SGW;][]{gott:05}, which has already been shown to strongly affect 
clustering \citep{zehavi:05,zehavi:10,nichol:06}.  We further show that a specific region 
dominates clustering measurements differently based on the galaxy sample and scale. 
No one structure dominates all scales, but almost all measurements are affected by 
at least one region.

\acknowledgments 

We are grateful for enlightening discussions and comments from many in the SDSS collaboration.  
We would like to specifically acknowledge valuable input from Istv\'an Szapudi, 
David H. Weinberg, Zheng Zheng, K. Simon Krughoff, Robert Nichol, Felipe Marin, Ravi Sheth, 
Robert E. Smith, Andrew Zentner, and Andreas A. Berlind.  We thank David Turnshek for his 
suggestions on improving the clarity of our descriptions.

We thank August Evrard and J\"{o}rg Colberg for kindly providing data and assistance with 
the Hubble Volume (HV) simulation.  The HV simulation was carried out by the Virgo 
Supercomputing Consortium using computers based at the Computing Centre of the Max-Planck 
Society in Garching and at the Edinburgh parallel Computing Centre.  

We are extremely appreciative of Michael Blanton for his work with the NYU-VAGC and the 
help he provided in our use of it.

%
%
%
%

I.~Z. acknowledges support by NSF grant AST-0907947.
J.~G. and the development of \ntropy\ was funded by NASA Advanced Information Systems Research Program grant NNG05GA60G.
A.~J.~C. acknowledges partial support from DOE grant DE-SC0002607 and NSF grant AST 0709394.

This research was supported in part by the National Science Foundation through 
TeraGrid resources provided by NCSA (Mercury) and the PSC (BigBen) under grant 
numbers TG-AST060027N and TG-AST060028N.


Funding for the SDSS and SDSS-II has been provided by the Alfred P. Sloan Foundation, 
the Participating Institutions, the National Science Foundation, the U.S. Department 
of Energy, the National Aeronautics and Space Administration, the Japanese 
Monbukagakusho, the Max Planck Society, and the Higher Education Funding Council for 
England. The SDSS Web Site is \texttt{http://www.sdss.org/}.

The SDSS is managed by the Astrophysical Research Consortium for the Participating 
Institutions. The Participating Institutions are the American Museum of Natural 
History, Astrophysical Institute Potsdam, University of Basel, University of 
Cambridge, Case Western Reserve University, University of Chicago, Drexel 
University, Fermilab, the Institute for Advanced Study, the Japan Participation 
Group, Johns Hopkins University, the Joint Institute for Nuclear Astrophysics, the 
Kavli Institute for Particle Astrophysics and Cosmology, the Korean Scientist 
Group, the Chinese Academy of Sciences (LAMOST), Los Alamos National Laboratory, 
the Max-Planck-Institute for Astronomy (MPIA), the Max-Planck-Institute for 
Astrophysics (MPA), New Mexico State University, Ohio State University, University 
of Pittsburgh, University of Portsmouth, Princeton University, the United States 
Naval Observatory, and the University of Washington.


\begin{appendix}

\section{Measurement Systematics}
\label{s:systematics}

\subsection{Estimating the Correlation Functions}
\label{ss:codetest}

A detailed comparison between 2PCF estimators by \citet{kerscher:00} found that 
$\widehat{\xi}_{LS}$ performs as well or better than other formulations. 
An investigation of three-point estimators shows $\widehat{\zeta}_{SS}$ performs 
favorably to alternatives, with stable estimates when using the least number of 
randoms \citep[see appendix in][]{kayo:04}.

The validity of our \npoint\ calculator is of paramount importance.  We carefully 
scrutinized the results of our main analysis code by separately implementing naive 
calculators for spatial and projected counts for both the 2PCF and 3PCF.  In addition 
to these internal consistency checks, we verified the accuracy with external codes 
used by different research groups.  For the spatial 3PCF, we reproduced exact 
results to those calculated by \texttt{npt} \citep{gray:04}.  We also checked that our 
projected 2PCF produced identical measurements to the code used by \citet{zehavi:05}.  

\subsection{Effects of Sky Completeness}
\label{ss:sky_comp}

Our measurements must take into account the angular sky completeness of the survey.  
We restrict our analysis to volume-limited galaxy samples, and as such, the radial selection
function typically used in flux-limited samples plays no role (it is defined to be unity for 
at all redshifts within the galaxy samples). The sky completeness can vary due to factors such 
as missing plates (specific regions in the sky), poor quality spectra and fiber collisions (see
\S\ref{s:sdss}).  The completeness is well characterized by sectors, and calculated by comparing 
the number of targeted galaxies with the corresponding number of spectra obtained \citep{vagc}.
We correct the estimated correlation functions by applying a multiplicative weight to pair counts 
such that each galaxy uses a weight assigned from the inverse completeness of the respective region. 
For the large scales, this weighting \emph{corrects} the clustering strength for regions where you
know galaxies exist, and is standard practice for measurements \citep{zehavi:02,zehavi:05,zehavi:10}.  
To test the magnitude of this correction, we consider a sample of $73\,320$ galaxies from an earlier
SDSS release \citep[DR5;][ specifically referred to as \texttt{safe26}]{sdss_dr5}, where we only 
consider galaxies with completeness between $0.5$ and $1.0$, with average and median values above $0.96$.

\begin{figure*}
  \centerline{
    \includegraphics[angle=0,width=0.9\textwidth]{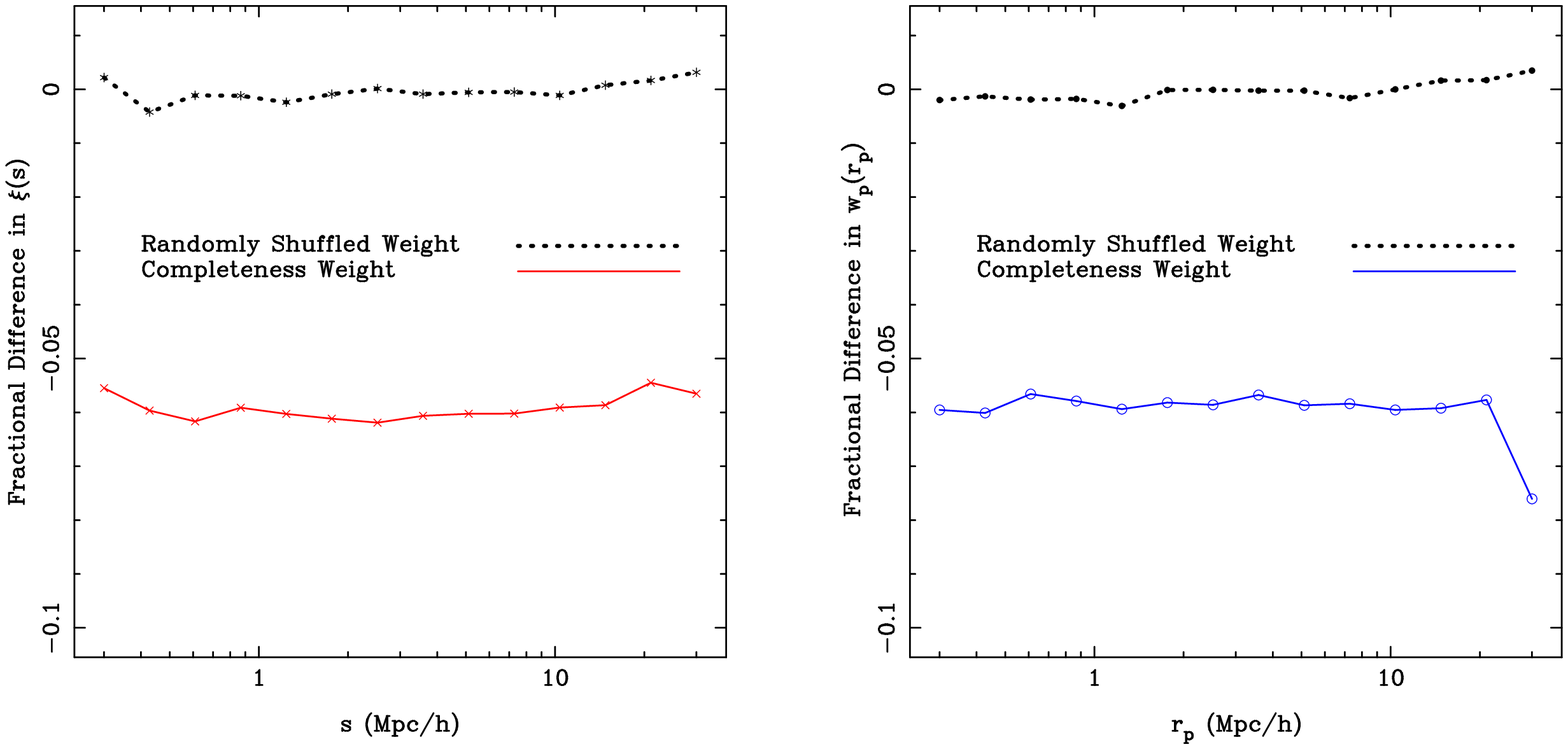}
  } 
  \caption[Effects of Sky Completeness on 2PCF]{
    We show the fractional difference when correcting for sky completeness in the 2PCF in redshift
    (left panel) and projected (right) space.  The fractional difference is defined as 
    $(\xi - \xi_{w}) / \xi_{w}$ where $\xi_{w}$ denotes the weighted (corrected) quantity.  
    We show two galaxy measurements using two weighting schemes, where we vary the weight 
    in calculating $\xi_{w}$. The solid lines depict weighting by the \emph{real} sky completeness, 
    and the dotted lines represent the same weights, but randomly assigned to galaxies. 
    As we expect, the randomly shuffled weights show little difference with the unweighted 2PCF.
  } \label{f:skyeffect_2pcf} 
\end{figure*}

We first show the effect of weighting by sky completeness on the 2PCF in
Figure~\ref{f:skyeffect_2pcf}.  We see this sample shows an approximately $6\%$ offset in both
redshift and projected space at all scales below $30 \hmpc$ if the weights are unaccounted for 
(see solid lines in both panels).  To demonstrate the validity of this test, we have re-measured the 
exact same galaxy data but randomly shuffled the weights which shows no fractional difference
(dotted lines).

\begin{figure*}
  \centerline{
    \includegraphics[angle=270,width=0.65\textwidth]{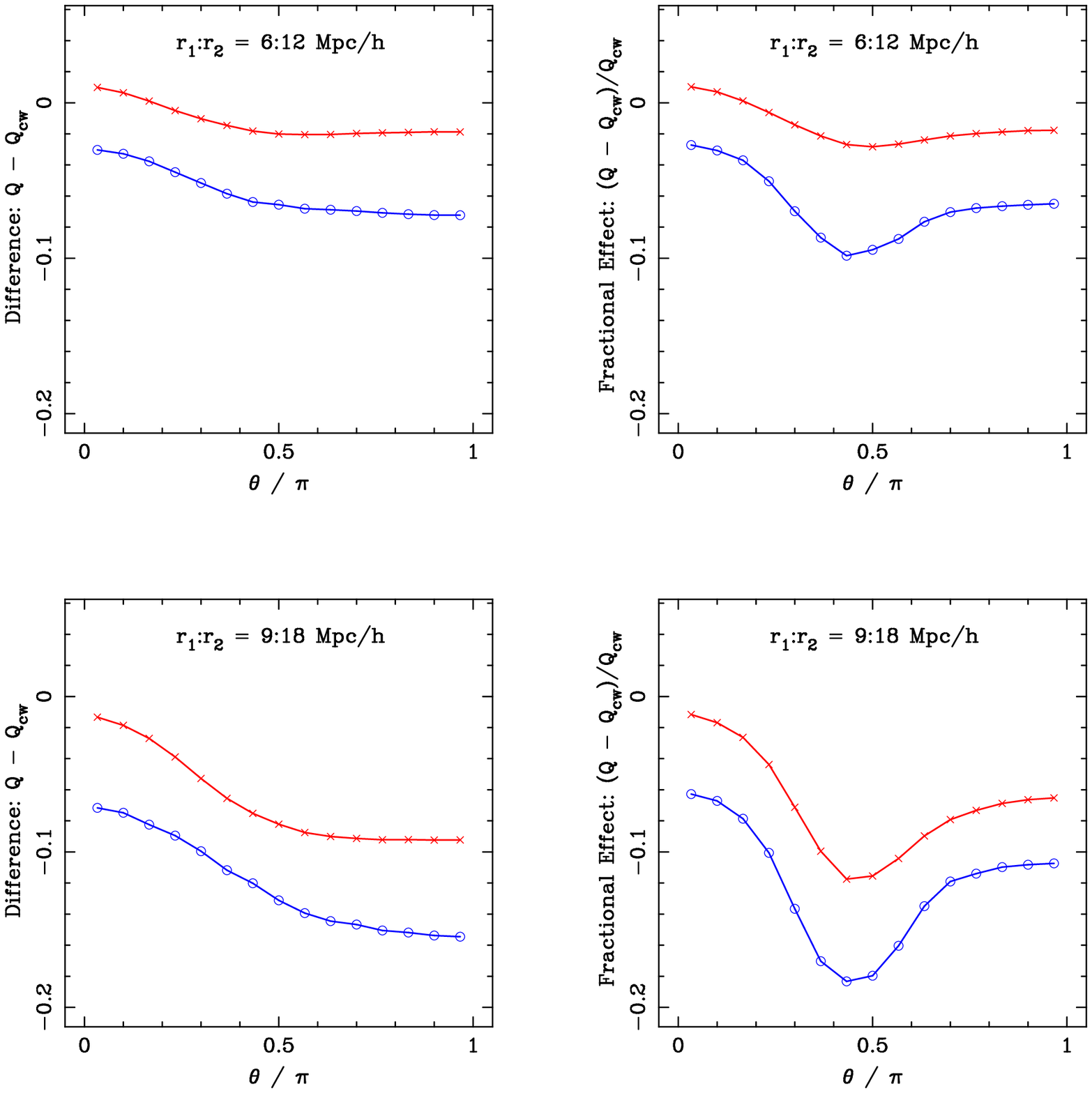} 
  } 
  \caption[Effects of Sky Completeness on $Q(\theta)$]{
    The effects of correcting for sky completeness in the reduced 3PCF.
    The $Q_{cw}$ denotes the completeness weighted (corrected) quantity.  The effect for 
    both redshift space (red, x-marks) and projected space (blue, open circles) measurements are 
    described by the absolute difference (left) and the fractional effect (right).  Not accounting 
    for sky completeness introduces a systematic offset that varies by scale, which can be seen 
    in the left panel.  The projected measurements appear more sensitive to sky completeness 
    than redshift space ones.
  } 
  \label{f:skyeffect} 
\end{figure*}

We investigate the effects of neglecting sky completeness on the reduced 3PCF in
Figure~\ref{f:skyeffect}, where we show both the absolute difference (left column) and fractional
effect (right).  We notice a systematic offset that increases with scale in both redshift and
projected measurements.  On larger scales ($r_1 = 9 \hmpc$), we find about a 12\% deviation in
redshift space and almost 20\% difference for projected space, showing more sensitivity to 
completeness weighting than the 2PCF.  We estimate the uncertainties in these measurements 
from $30$ jackknife samples at approximately 12\% and 17\% respectively, and we conclude that 
sky completeness must be considered for accurate 3PCF measurements.

\subsection{Effects of Binning}
\label{ss:binning}

The choice of binning scheme affects measurements and covariance matrices of the 
configuration dependence in the 3PCF.  To be specific, we refer to the choice of bin-size as 
well as their spacing by ``binning scheme''.  For the 2PCF, the most common scheme is to use 
log spaced bins in $r$ to account for the dynamic range of scales and power law dependence 
of the 2PCF.  In this scheme, bins at larger $r$ correspond to larger bin-widths.
When we measure $Q(\theta)$, the angular bins are closely packed over a much smaller scale 
range making a choice of log spaced bins impractical. 
If the bin-size is too large, we notice an attenuation or averaging out of the configuration 
dependence.  On the other hand, if the bin-size is too small, we do not resolve the measurement 
nor the covariance.  Both the sample size and number density will impact the efficiency 
of a binning scheme. Even if we find an acceptable choice for one sample, it might show 
different effects on another sample.  This effect remains more dramatic for the 3PCF due to 
the larger parameter space of potential configurations.

Naively, for a given data set of size $N$, we expect the number of available triplets (i.e. $N^3$) 
to be much larger than pairs ($N^2$).  However, the configuration dependence of the 3PCF is a 
function of three dependent variables. Any specific configuration represents a tiny slice through 
the available data, making the 3PCF much harder to determine.  For a rough idea, we can compare the 
number of triplets to the number of pairs for a specific 3PCF configuration.  For a choice of 
binning scheme with \emph{no} overlap for the $r_1 = 9 \hmpc$ triangles, we find that approximately 
$1$ in every $1000$ pairs contribute to the 2PCF for the denominator of $Q_{proj}(\theta)$.  
For the 3PCF (the numerator), only $1$ in $67,000,000$ contribute.  Even if we consider that the 3PCF 
scales as the square of the 2PCF (i.e. $\zeta \propto \xi^2$), we notice the triplet count in 
the 3PCF remains smaller by a factor of $\sim 67$.

The impact of bin-size has only recently been addressed in the literature \citep[see][]{GS05}.
We believe this remained unresolved for two main reasons (1) the availability of large datasets to 
statistically determine finely binned higher order moments, and (2) the computational complexity 
of calculating them. For a small data sample, it may not be possible to measure the 3PCF with small 
enough bins. This is not the case with our SDSS samples. Most methods of estimation solve the 
computational complexity by performing counts after pre-gridding the data, thereby imposing a 
bin-size effect.  While this can help mask the effect, it does not hide it entirely. 
\citet{GS05} comment on the effect of bin-size using a pre-gridding technique by using a sufficiently 
fine grid.  Our estimation of correlation functions uses an efficient counting algorithm to yield exact 
counts, so we do not need to pre-grid our data.  At large scales, measurements using a sufficiently 
fine pre-gridding technique and those with an exact count converge.  At small scales, pre-gridding 
becomes less effective at reducing the computational expense and may become prohibitively expensive.

\begin{figure*}
  \centerline{ 
    \includegraphics[angle=0,width=0.50\textwidth]{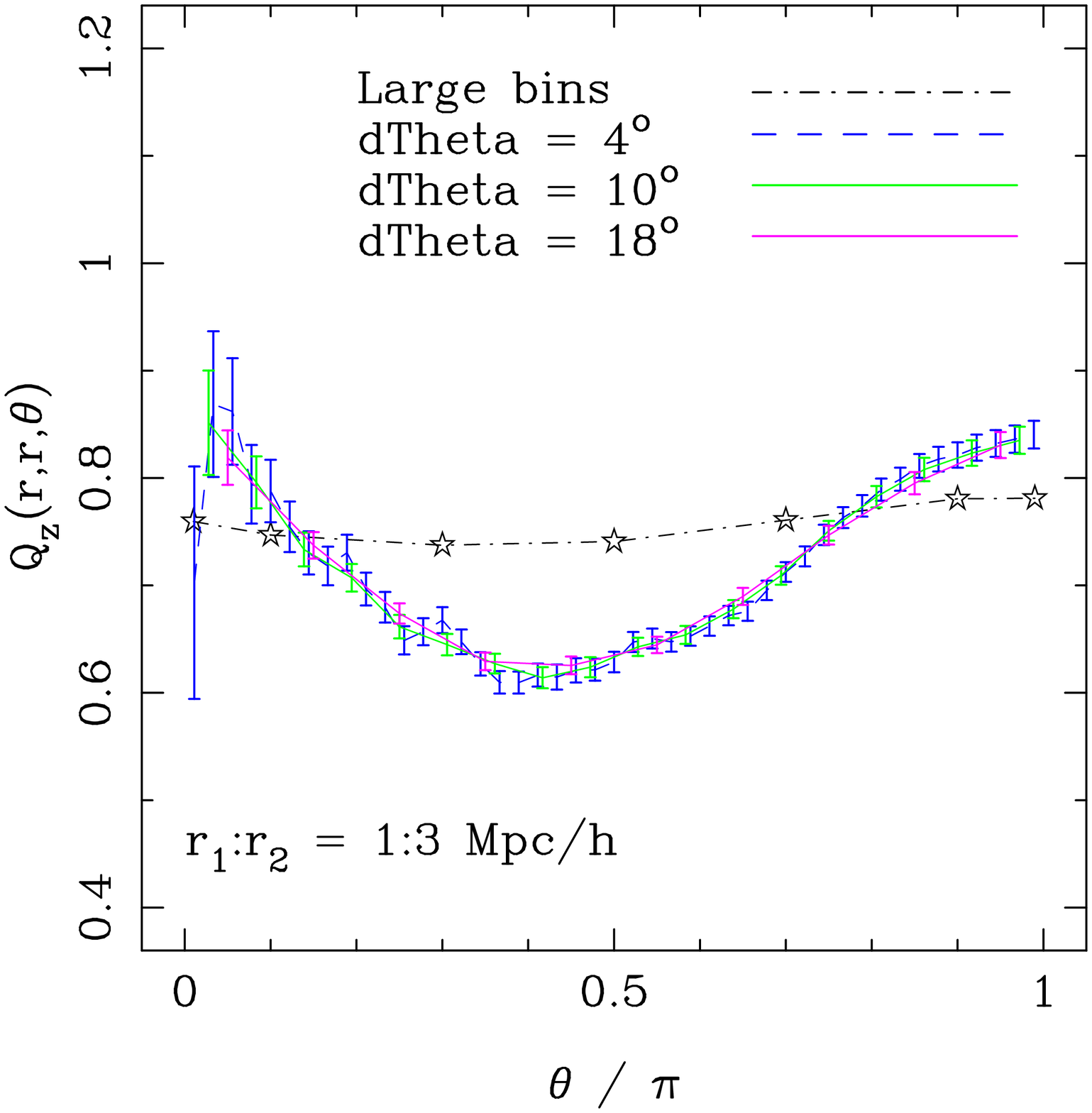}
  }
  \caption[Bin-size effects on small scale ``U-shape'' in $Q_z(\theta)$]{
    We show the reduced 3PCF on a subset of SDSS data.  At small scales, we expect the redshift
    space distortions to cause a ``U-shape'' signal due to elongated fingers-of-god.  This plot
    illustrates that small binning ($\pm 0.1 \hmpc$) in the two sides ($r_1$ and $r_2$) resolves the
    ``U-shape'' even with large $\theta$ bins.  For comparison, we show ``Large Bins'' defined to
    correspond to the binning scheme of \citet{nichol:06} (i.e. $\pm 0.5 \hmpc$).  The error bars
    show the combined Poisson errors of the respective counts and signify the statistical
    significance of the bin-sizes.
  } 
  \label{f:Q_bincheck} 
\end{figure*}

\citet{GS05} suggest a good test of bin-size is to measure the small scale configuration dependence 
of the reduced 3PCF in redshift space.  Redshift distortions produce elongated finger-of-god structures, 
which should significantly amplify the signal of collapsed triangles at $\theta = 0$ and $\theta = \pi$.  
This characteristic ``U-shape'' will not be present if the bin-size is too large, since the 
finger-of-god structures are effectively averaged over.  For the 3PCF, we must consider the bin-size 
in three parameters, i.e. each side of the triplet.  In figure \ref{f:Q_bincheck}, we show three choices 
of bin-size in $\theta$ when two sides of the triangle ($r_1,r_2$) are tightly constrained 
($\pm 0.1 \hmpc$).  Even with large bins in $\theta$, all measurements equivalently show 
configuration dependence, demonstrating the importance of bin-size for $r_1$ and $r_2$.  We also 
show a measurement with ``large bins'', constructed to match those used in \citet{nichol:06} where 
the first side of the triangle ($r_1$) varies by $\pm0.5 \hmpc$.  The larger bin-size completely 
masks the configuration dependence expected from the redshift distortions.  

\begin{figure*}
  \centerline{ 
    \includegraphics[angle=270,width=0.95\textwidth]{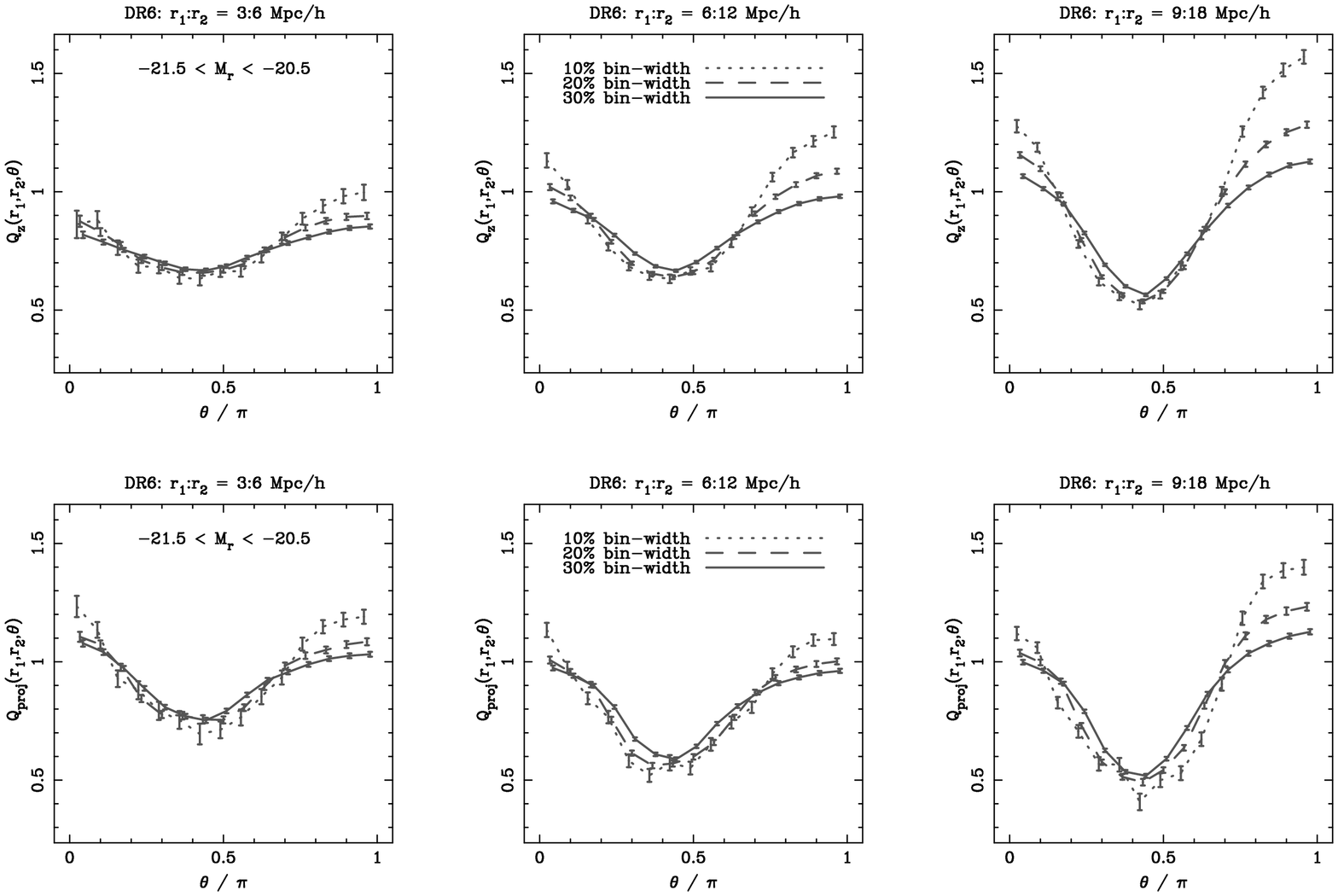}
  }
  \caption[Binning effects on $Q(\theta)$]{
    In our measurements on DR6 data, we adopt a fiducial binning scheme using linear spaced bins in 
    $\theta$ with a bin-size a set fraction of the midpoint. For the three triangle scales of interest, 
    we compare this scheme using three fractional bin-widths on 
    $Q_z$ (top row) and $Q_{proj}$ (bottom row): $10$, $20$ and $30$\%.  The error bars 
    represent Poisson errors on the counts from each bin.  Larger bin-width measurements show smaller 
    uncertainties and less configuration dependence.  
    We use the LSTAR galaxy sample from DR6 selected to have $-21.5 < M_r < -20.5$.
  } 
  \label{f:Q_bincmp} 
\end{figure*}

In Figure~\ref{f:Q_bincmp} we extend the comparison of bin-width to the three specific scales that we 
measure in redshift and projected space.  We utilize our \emph{fiducial} binning scheme, which
consists of 15 linear spaced bins in $\theta$ with the bin-width chosen to be a fraction (denoted with $f$)
of the scale of the bin midpoint.  To be clear, the bin-size for $r_1$ and $r_2$ also change appropriately with $f$. 
We show results for $f = 0.1, 0.2$ and $0.3$ with $1\sigma$ Poisson uncertainties calculated from bin counts.  
The larger bin-width smooths the configuration dependence at all scales, with a dramatic effect on the 
$r_1 = 9 \hmpc$ triangles.  This occurs in both redshift and projected space measurements of $Q(\theta)$.  
Physically, larger bins allow a greater range of configurations to be represented in each bin. 

\begin{figure*}
  \centerline{ 
    \includegraphics[angle=270,width=0.75\textwidth]{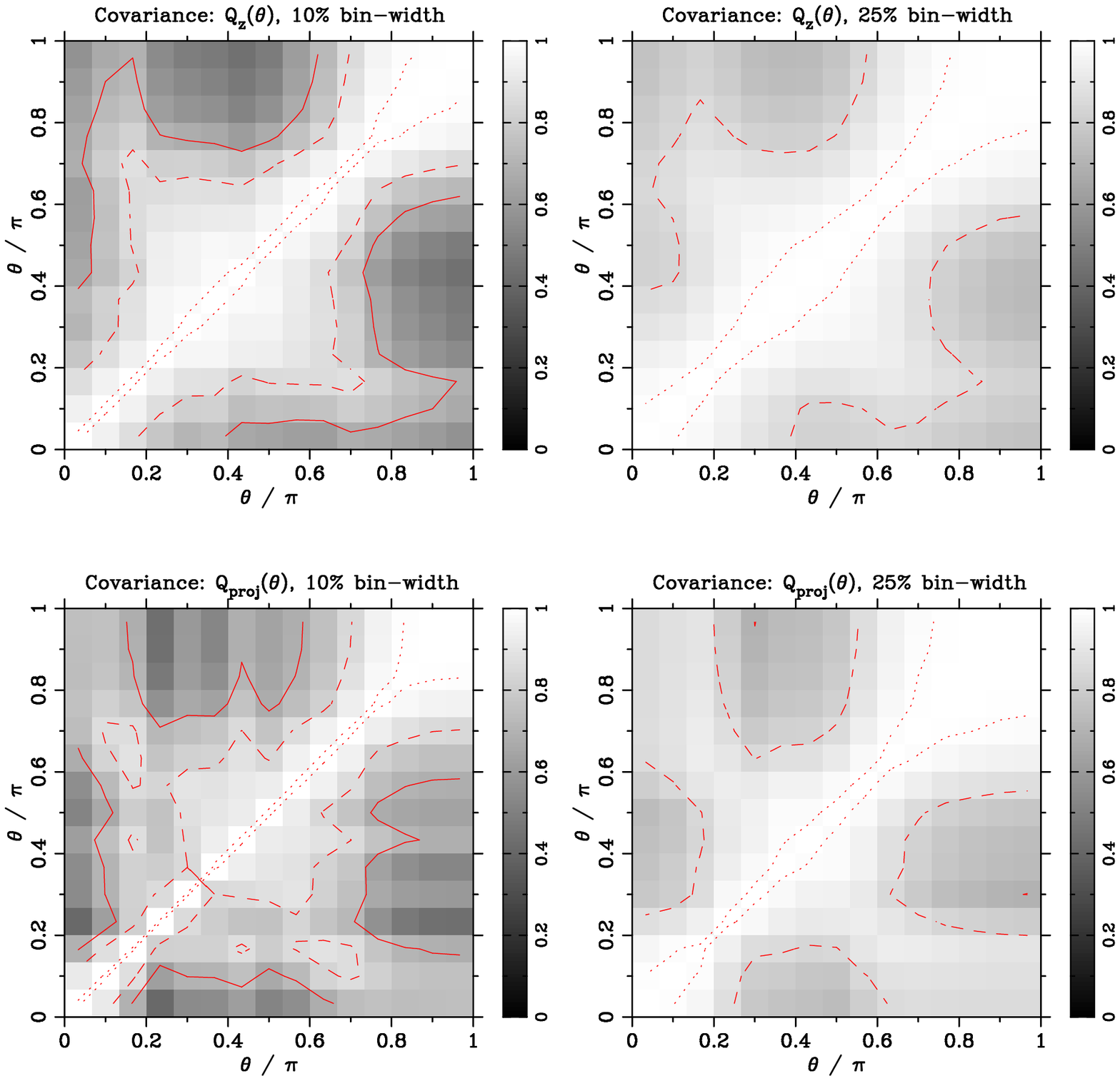}
  }
  \caption[Covariance matrices for $Q_z(\theta)$ and $Q_proj(\theta)$ with $f = 0.1$ \& $0.25$]{
    For our fiducial binning scheme, we show the covariance matrices for $Q_z(\theta)$ (top row) and 
    $Q_{proj}(\theta)$ (bottom row) for the largest triangles with $r_1 = 9 \hmpc$. 
    The solid, dashed 
    and dotted contours correspond respectively to values of 0.70, 0.85 and 0.99.
    We use the LSTAR galaxy sample from DR6 ($-21.5 < M_r < -20.5$).
  } 
  \label{f:Q_covar_fcmp} 
\end{figure*}

We keep the number of bins fixed (at 15) but vary the bin-width; this results in an increased 
overlap of configurations, and hence imposes a larger correlation in the covariance between 
neighboring bins.  Basically, the physical overlap is simply larger for equally spaced $\theta$ bins 
when $\theta$ corresponds to $\sim 0$ or $\sim\pi$.  We show the covariance matrices for $Q_z(\theta)$ 
and $Q_{proj}(\theta)$ in Figure~\ref{f:Q_covar_fcmp} for the largest scale ($r_1 = 9 \hmpc$) and two bin-widths 
using $f = 0.1$ and $0.25$.  The significant configuration overlap in $f=0.25$ results in a larger correlation, 
as we expect.  However, we also see increased correlation in non-overlapping bins 
(see the $\theta \sim 0$ with $\theta \sim \pi$ bins; the top left and bottom right corners).

Clearly identified differences due to bin-size are prevalent in both measurements of the reduced 3PCF and 
its covariance.  The general rule of thumb is that smaller bins, assuming they are well resolved, 
produce more accurate results -- but that assumes we properly resolve the covariance.  

\section{Projected Correlation Functions}
\label{s:proj}

The purpose of employing projected correlation functions is to minimize the effects of redshift 
distortions on clustering measurements. However, we might ask how effective the projected 
statistic really is.  We use the Hubble volume (HV) \nbody\ simulations to investigate this question.
Since we want to apply our results to observational galaxy surveys, we construct realizations 
of dark matter (DM) that have the same footprint and volume of an observation galaxy sample. 
This should incorporate any edge effects or issues of finite volume that are also present 
in galaxy data. We use line-of-sight peculiar velocities to create a redshift space 
distribution of DM.  In a few cases, we compare this to a similar distribution without any 
redshift distortions, which we call \emph{real} space.  Two important decisions need to be 
made when using projected measurements: (1) how large of $\Delta\pi$ bins do we use to 
integrate over and (2) what is the maximum line-of-sight distance for the integration 
($\pimax$).  

\begin{figure}
  \plottwo{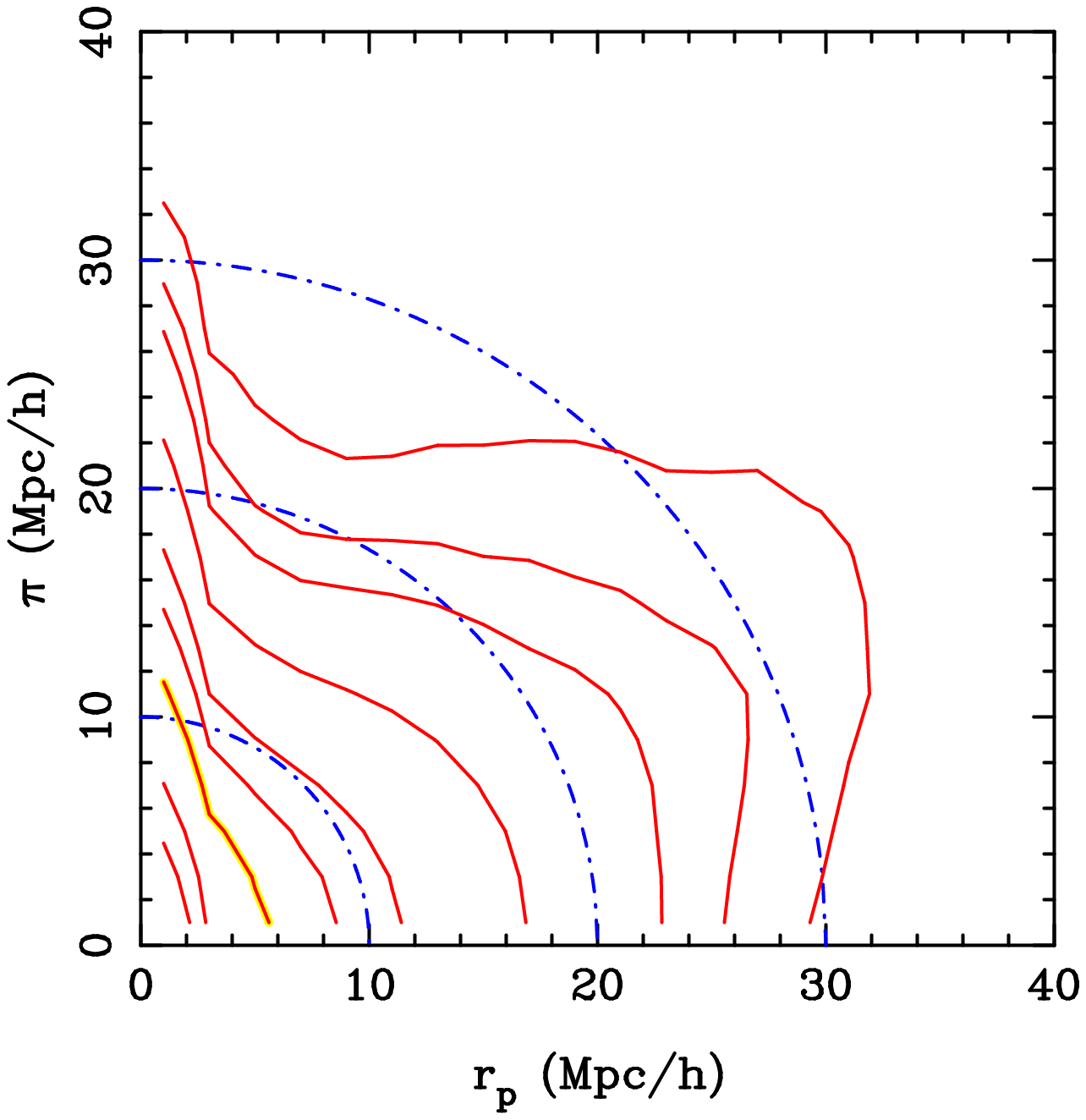}{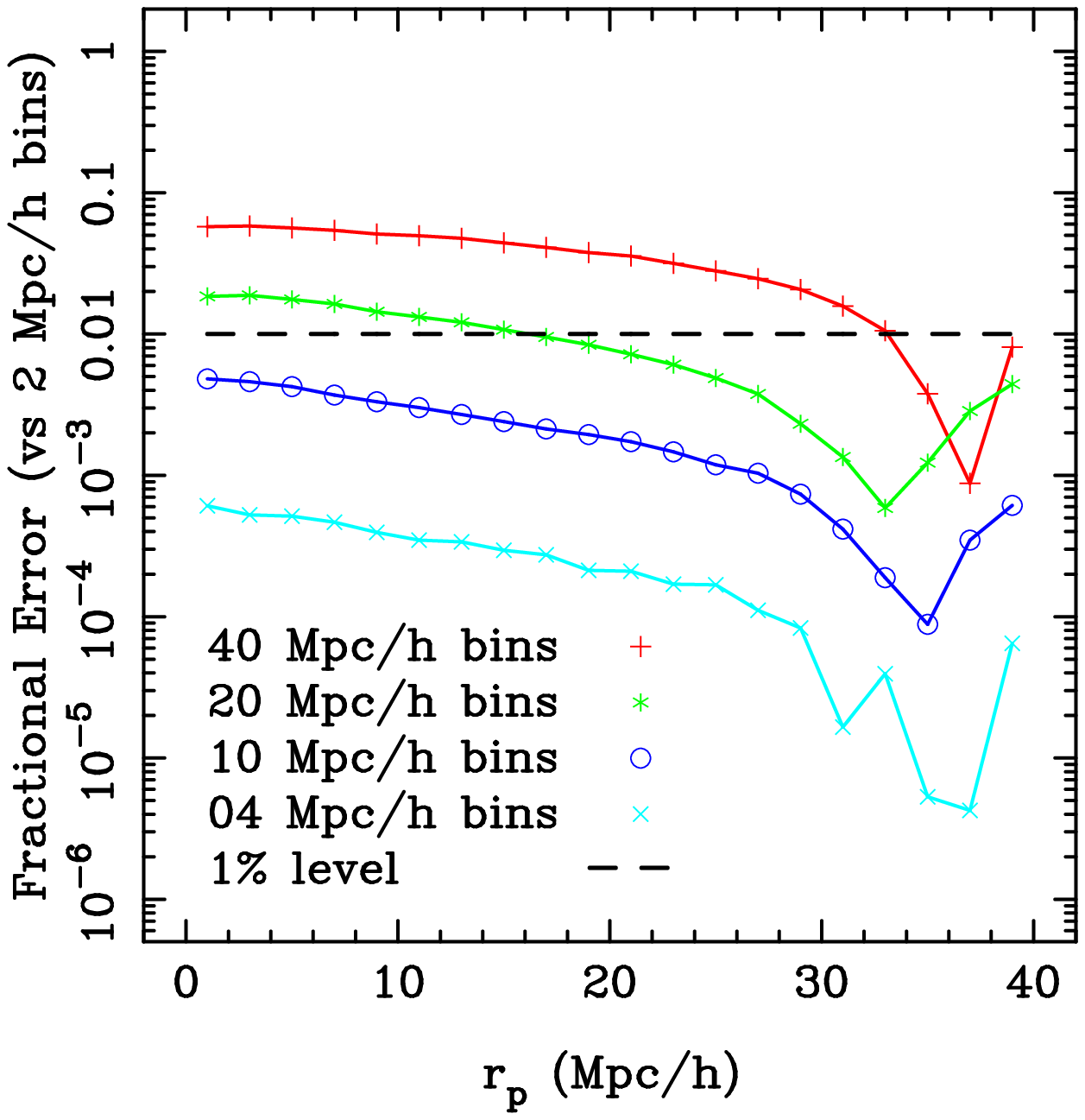} 
  \caption[HV simulation: $r_p - \pi$ diagram and effects of $\pi$ bin-width]{
    On the left we show the $r_p - \pi$ diagram from the 2PCF on dark matter particles from the Hubble Volume 
    simulation, using a grid of $2 \hmpc$ in both $r_p$ and $\pi$.  
    The HV particles are trimmed to 
    match the same volume and footprint of SDSS data, with velocities used to construct 
    radial redshift distortions.  The red (solid) lines are contours of a specific value 
    of the CF, $\xi(r_p,\pi)$, with values of $3,2,1, 0.6, 0.4, 0.2, 
    0.1, 0.07, 0.04$ with the yellow highlighted one corresponding to 
    $\xi(r_p, \pi) = 1$ . The blue (dot-dashed) semi-circles show a perfect
    isotropic correlation for comparison.
    On the right, we depict the accuracy of the $w_p(r_p)$ integration using different sized $\pi$ bins, 
    which we show as a function of projected scale ($r_p$). We obtain the ``truth'' value by comparing to a 
    fiducial $\Delta\pi = 2 \hmpc$, and our $\pimax = 40 \hmpc$.  We show the 1\% level of 
    accuracy with the dashed line.
  } \label{f:hv_wptest} 
\end{figure}

The first question stated above is relatively easy to address.  We want to choose bins of 
$\Delta\pi$ that remain small enough to prevent a smoothing bias.  In the left panel of 
Figure~\ref{f:hv_wptest}, we show the anisotropy introduced by redshift distortions on the 2PCF, 
the $r_p - \pi$ diagram, within our HV test sample. Using these data, we calculate the projected 
2PCF by integrating to $\pimax = 40 \hmpc$ with different values of $\Delta\pi$.  We compare 
each with the fiducial measurement using $\Delta\pi = 2 \hmpc$, and show the fractional difference in 
Figure~\ref{f:hv_wptest}.  We notice a very small effect, and the largest deviation appears 
at small projected separation where $\xi(r_p,\pi)$ changes rapidly.
The systematic effect remains below the 2\% level even with bins as wide as $\Delta\pi = 20 \hmpc$.

\begin{figure*}[ht]
  \centerline{
    \includegraphics[angle=0,width=0.65\textwidth]{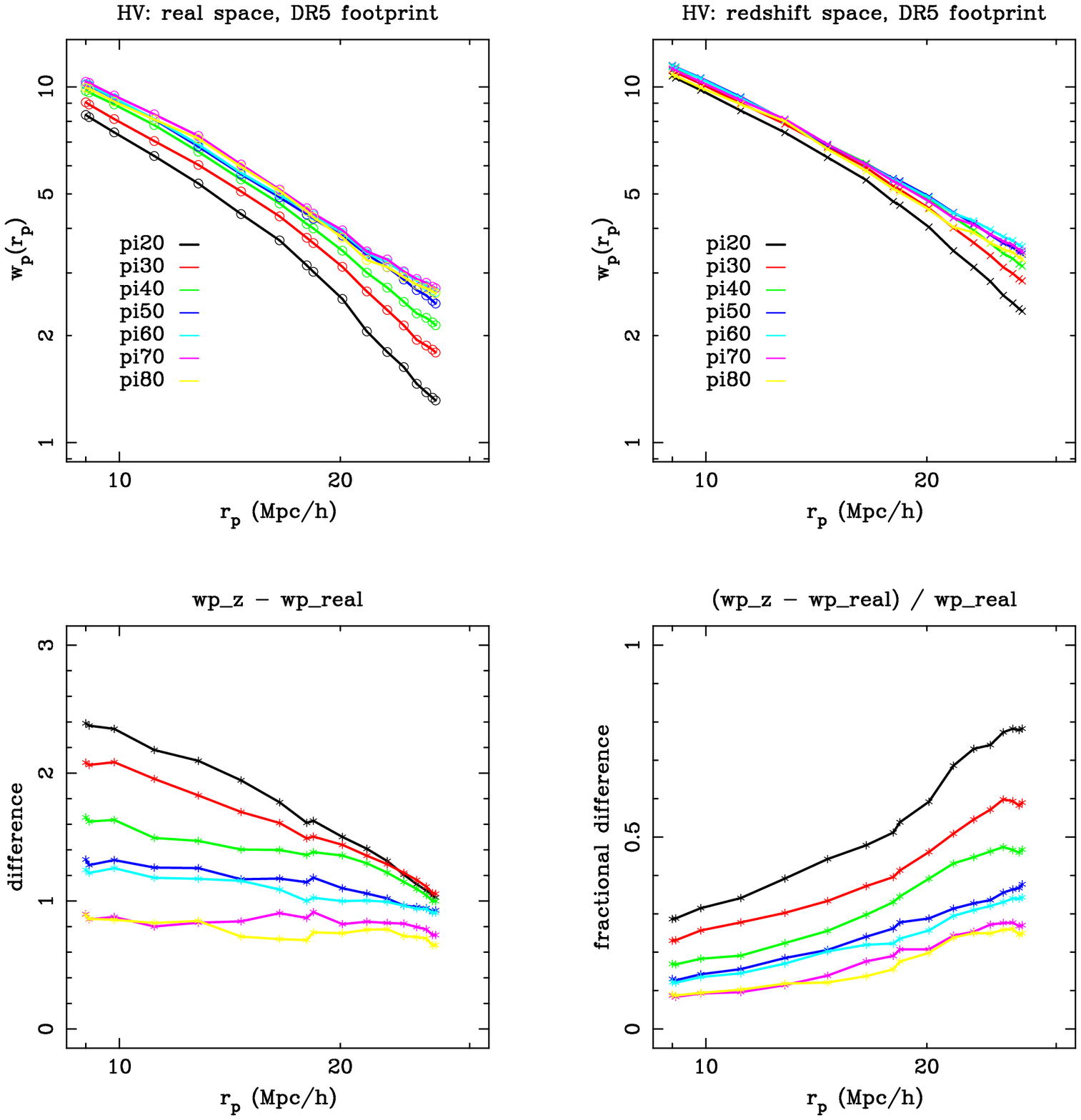}
  }
  \caption[The Effects of $\pimax$ on $w_p(r_p)$ in simulations]{
    We show the projected 2PCF, $w_p(r_p)$, for dark matter particles in the Hubble Volume 
    simulation both in real space (top left panel), redshift space (top right panel), 
    the absolute difference due to redshift distortions (bottom left panel), and the fractional 
    difference (bottom right panel). The different curves correspond to different $\pimax$ 
    integrations, as denoted.  We estimate $w_p(r_p)$ by integrating in bins of 
    $\Delta\pi = 5 \hmpc$.  The legend in the top two panels denotes the maximum 
    $\pimax$ used for each calculation, such that ``pi20'' corresponds to $\pimax = 20 \hmpc$, 
    etc. 
  } \label{f:hv_pitest}
\end{figure*}

The second question remains more subtle: what is an appropriate $\pimax$?  Formally, 
we might prefer to keep $\pimax = \infty$, which preserves the property that a power law spatial
correlation function $\xi$ produces a power law projected 2PCF $w_p$.  Realistically, galaxy samples 
represent finite volumes and the correlation function can only be well estimated to a certain
maximum distance.  We want to keep our $\pimax$ well below this limit, lest we dilute our signal.
On the other hand, we want $\pimax$ be large enough to (re-)capture clustering strength lost due 
to redshift distortions.  We compute $w_p(r_p)$ using two same volume realizations of DM, one in real 
space (no distortions) and one in redshift space, which we compare in Figure~\ref{f:hv_pitest}.  
We expand $\pimax$ to range between $20$ and $80 \hmpc$.  We see several interesting features.  
First, both real and redshift space $w_p(r_p)$ are roughly power laws, with the strongest power
corresponding to the largest $\pimax$.  We expected this, as the projected 2PCF has units 
of distance, a larger integration equates to a higher functional value.  We notice a much smaller 
difference between the smallest and largest $\pimax$ in redshift space.  This 
is a result of the large scale infall \citep{kaiser:87}, as clustering strength is compressed to
smaller line-of-sight separation which we see in Figure~\ref{f:hv_wptest}.  Examining the fractional 
difference between real and redshift space measurements, we find a higher $\pimax$ does what we 
expect in reducing the difference between real and redshift space $w_p(r_p)$ measurements.  
However, we also see that different values of $\pimax$ dramatically changes the 2PCF from 
the ``ideal case'', inducing a systematic difference which can range from right below the 10\% 
level at $r_p = 10; \pimax = 80\hmpc$ to 80\% at the largest scales and smallest $\pimax$.
Finally, we conclude that even at $\pimax = 80\hmpc$ the effects of redshift distortions cannot 
be completely negated by the projected 2PCF.

\begin{figure*}[hb]
  \centerline{
    \includegraphics[angle=270,width=0.9\textwidth]{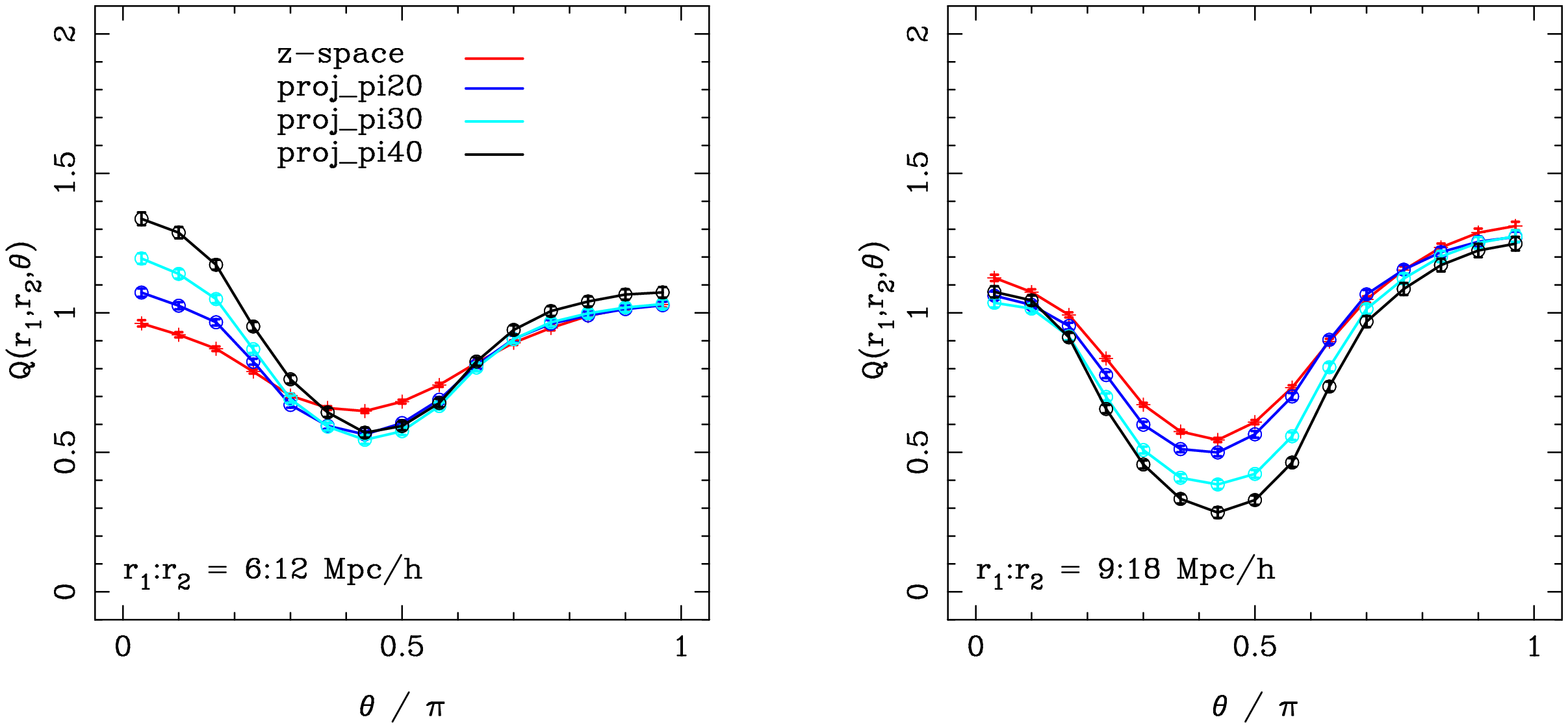}
  }
  \caption[Effects of $\pimax$ on $Q_{proj}$ in observed galaxies]{
    We show the reduced 3PCF, $Q(\theta)$, on a galaxy sample from SDSS DR5 at our two largest
    triangle scales: $r_1 = 6 $ \& $9 \hmpc$.  We compare three different values of $\pimax$, 
    namely $20,30$ and $40 \hmpc$ with the \emph{spatial} reduced 3PCF in redshift space.  
    We note the lower values of $\pimax$ tend toward the redshift space measurement, but 
    emphasize that it is a minor effect at these scales.
  } \label{f:Qprojtest} 
\end{figure*}

We extend the investigation to the reduced 3PCF, using the two largest triangle configurations
($r_1 = 6$ and $9 \hmpc$).  Since $Q(\theta)$ is normalized by the respective 2PCF, we expect the
same amplitude between spatial and projected measurements and can compare measurements on real
observed galaxy samples.  We show the measurements on an SDSS sample where we measure $Q(\theta)$ in 
redshift space and projected with $\pimax = 20,$ $30$ and $40\hmpc$ in Figure~\ref{f:Qprojtest}.  
For observational galaxies, it appears that $Q_{proj}$ recovers some configuration dependence with 
larger $\pimax$.  The $r_1 = 9 \hmpc$ triangles show decreased amplitude perpendicular configurations 
and we note the ``splitting'' of collapsed triangles ($\theta \sim 0$) in the smaller triangles.  In 
both cases, a decreased $\pimax$ yields a $Q(\theta)$ that approaches the redshift space measurement, 
in agreement with our expectations.

Overall, we conclude that the projected correlation function reduces the impact of 
redshift distortions on measurements of clustering.  However, the projected statistic does 
not completely remove effects of the distortions for any of practical values of $\pimax$ 
that we might use.  Our results suggest a larger $\pimax$ might further minimize redshift 
distortions but a more thorough investigation is required to disentangle the systematics 
(for example, we integrate the projected 3PCF with a single bin of $\pimax$ 
width which we would have to investigate).  Since the computational efficiency of 
estimating the 3PCF decreases dramatically with increased $\pimax$, we choose to use 
$\pimax = 20 \hmpc$ for our measurements. This choice should be sufficient to minimize some 
effects of redshift distortions, remaining most effective at small scales where it will 
recapture clustering strength lost from the non-linear collapse in dense regions 
(the fingers-of-god).

\end{appendix}

\bibliography{refs}

\end{document}